\documentclass[draft,tightenlines,nofootinbib,preprint,aps,eqsecnum]{revtex4}

\usepackage{latexsym}

\newcommand{\beq}{\begin{equation}}
\newcommand{\eeq}{\end{equation}}
\newcommand{\bea}{\begin{eqnarray}}
\newcommand{\eea}{\end{eqnarray}}
\newcommand{\cir}{{\buildrel \circ \over =}}

\begin{document}

\title{Dirac-Bergmann Constraints in Physics: Singular Lagrangians, Hamiltonian Constraints and the Second Noether Theorem }

\medskip

\author{Luca Lusanna}

\affiliation{ Sezione INFN di Firenze\\ Polo Scientifico\\ Via Sansone 1\\
50019 Sesto Fiorentino (FI), Italy\\  E-mail: lusanna@fi.infn.it}

\today

\begin{abstract}

There is a review of the main mathematical properties of system described by singular Lagrangians and requiring Dirac-Bergmann 
theory of constraints at the Hamiltonian level. The following aspects are discussed:\medskip

i) the connection of the rank and eigenvalues of the Hessian matrix in the Eulerr-Lagrange equationsù
with the chains of first and second class constraints;

\medskip

ii) the connection of the Noether identities of the second Noether theorem with the Hamiltonian constraints;

\medskip

iii) the Shanmugadhasan canonical transformation for the identification of the gauge variables and for the search of the Dirac observables,
i.e. the quantities invariant under Hamiltonian gauge transformations.

\bigskip

Review paper for a chapter of a future book.

\end{abstract}

\bigskip

\maketitle

\section{Introduction}

Most of the relevant interactions in physics are described by singular Lagrangians implying the presence of Dirac-Bergmann constraints \cite{1,2,3}
at the Hamiltonian level. This happens for electro-magnetism, for the standard model of particle physics ($SU(3) \times SU(2) \times U(1)$ Yang-Mills fields)
and its extensions, for Einstein theory of gravity and for all its generally covariant variants. Also the description of relativistic classical and quantum point particles, needed for bound states in the particle approximation of quantum field theory (QFT), requires Hamiltonian constraints for the elimination of relative times (no time-like excitation is seen in spectroscopy;
see the review in Ref.\cite{4} and its bibliography).

\medskip

In all these theories the main problem at the classical level is the identification of the gauge-invariant physical degrees of freedom, the so called Dirac observables (DO). Instead
the main open problem at the quantum level is whether one has to quantize only the DO's or also the gauge variables shifting the search of the physical observables after
quantization like in the BRST approach.

\bigskip

In this Chapter I will present a review of the main properties of constrained systems  based on my personal viewpoint on the subject at the classical level with some comments
on the weak points of the existing quantization approaches. Then in another Chapter  I will show their use in special relativity, gauge theories and gravity.
\medskip

Besides Dirac's book \cite{1} and Ref.\cite{5} I recommend the books in Refs.\cite{6,7} for an extended treatment of many aspects of the theory also at the quantum level
(included the BRST approach). Other books on the subject are in Refs. \cite{8,9,10,11}. Instead there is no good treatment of constrained
systems in mathematical physics and differential geometry: there are only partial treatments for finite-dimensional systems like presymplectic geometry \cite{12,13}
(see Refs.\cite{14,15,16,17} and their bibliography for recent contributions) without any extension to infinite-dimensional systems like field theory \cite{18}.

\bigskip

In Section II there is a short review of regular Lagrangian systems for finite-dimensional systems and of their description in the Hamiltonian and velocity space formalisms.

In Section III there is the definition of singular Lagrangians and the description of the Hamiltonian constraints arising in phase space. After the introduction of the Dirac multipliers
there is the formulation of the Dirac algorithm for finding the final constraint manifold. After the definition of first and second class constraints there is the diagonalization of the Dirac algorithm. Then the notion of Dirac brackets is introduced for the determination of a phase space without second class constraints.

In Section IV there is the study of the Hamiltonian gauge transformations generated by the first-class constraints. Then the gauge invariant quantities, named Dirac observables (DO), are defined and the reduced phase space is defined. Then the Shanmughadasan canonical transformations for the determination of canonical bases containing a set of DO's for the physical degrees of freedom are described.

In Section V there is the study of the eigenvalues of the Hessian matrix and of the Euler-Lagrange (EL) equations when the rank of the Hessian matrix is constant. Then there is a sketch of the pathologies which can appear when such a rank is not constant: proliferation of constraints, ramification of constraint chains, third and fourth class constraints...

In Section VI, after a review of the first Noether theorem and of its extensions there is the description of the second Noether theorem for singular Lagrangians. It is shown which is the connection of the resulting Noether identities with the constraints of the Dirac algorithm.

In Section VII there is the extension to field theory of constraint theory.

Some open problems are described in the Conclusions.

\section{Regular Lagrangians for Finite-Dimensional Systems}

Let us consider a finite-dimensional system whose {\it
configuration space} $Q$ is $n$-dimensional (either $Q = R^n$ or $Q$ is an
$n$-dimensional manifold with or without boundary), spanned by
the configurational  coordinates $q^i$, $i=1,..,n$.
We shall give a short review of standard classical mechanics for such systems \cite{19,20,21,22}.

\subsection{The Second Order Lagrangian Formalism.}

Let the system be described by a time-independent {\it Lagrangian}
$L(q(t), \dot q(t))$, where $q^i(t)$ is a curve in $Q$ with time
as parameter and ${\dot q}^i(t) = {{d q^i(t)}\over {d t}}$ are the
velocities, and by the Lagrangian action $S = \int^{t_f}_{t_i} dt\, L(q,\dot q)$.

The stationarity of the action, $\delta S = \int dt\,
{{\delta S}\over {\delta q^i(t)}}\, \delta q^i(t) =0$, under variations
$\delta q^i(t)$ [$\delta {\dot q}^i = {d\over {d t}} \delta q^i$]
which vanish at the endpoints $t_i$, $t_f$, identifies the
classical motions of the system as those trajectories $q^i(t)$
which satisfy the Euler-Lagrange (EL) equations (the
summation convention over repeated indices is used;
the symbol $\cir$  denotes an equality which holds
only on the trajectories solution of the equations of motion)

 \bea
  L_i &=& {{\partial L}\over {\partial q^i}} - {d\over {d t}}\,
  {{\partial L}\over {\partial {\dot q}^i}} =- ( A_{ij} {\ddot
  q}^j - \alpha_i) \cir 0,\nonumber \\
   &&\alpha_i(q,\dot q) = {{\partial L}\over {\partial q^i}} - {{\partial^2
 L}\over {{\dot q}^i\, \partial q^j}}\, {\dot q}^j = (1- {\dot q}^k\, {{\partial}\over {\partial {\dot q}^k}})\,
{{\partial L}\over {\partial q^i}} - R_{ij}\, {\dot q}^j,\nonumber \\
&&R_{ij}(q,\dot q) = - R_{ji} = {{\partial^2 L}\over {\partial {\dot q}^i\,
  \partial q^j}} - {{\partial^2 L}\over {\partial {\dot q}^j\,
  \partial q^i}}.
 \label{2.1}
 \eea

The {\it Hessian} matrix is $A_{ij}(q,\dot q) = A_{ji} = {{\partial^2 L}\over {\partial {\dot q}^i\,
 \partial {\dot q}^j}}$ and the Lagrangian is said {\it regular} when $det\, A
\not= 0$. If we denote $B = A^{-1}$ the inverse Hessian matrix, it
follows that the EL equations can be put in the following {\it
normal} form ${\ddot q}^i -\Lambda^i \cir 0$, $\Lambda^i = B^{ij}\,
  \alpha_j$.

\subsection{The First Order Hamiltonian Formalism.}

The {\it canonical momenta} are defined by $p_i = {{\partial L(q, \dot q)}\over {\partial {\dot q}^i}} = {\cal P}_i(q, \dot q)$
and the regularity condition $det\, A \not= 0$ implies
that these equations can be inverted to express the velocities
${\dot q}^i$ in terms of $q^k$ and $p_k$.

By means of the Legendre transformation we can re-formulate the
second order Lagrangian formalism in the first order Hamiltonian
one on the phase space $T^*Q$ (the co-tangent bundle) over $Q$
with coordinates $q^i$, $p_i$. The Hamiltonian of the system
is $\bar H = p_i\, {\dot q}^i - L$ (we shall denote $\bar f =\bar f(q,p)$ the functions on
phase space) and the phase space action is $\bar S = \int^{t_f}_{t_i} dt\, \bar L$ with $ \bar L = p_i\,
{\dot q}^i - \bar H$. By asking the stationarity, $\delta \bar S =0$, of this action
under variations ${\bar {\delta q^i}}$ which vanish at the
endpoints $t_i$, $t_f$, and under arbitrary variations ${\bar
{\delta p_i}}$, we get the first order differential Hamilton
equations of motions ${\bar {L_{qi}}} = {\dot q}^i - {{\partial \bar H}\over
{\partial p_i}} \cir 0$, ${\bar {L_{pi}}} = {\dot p}_i + {{\partial \bar H}\over {\partial
q^i}} \cir 0$. The first half of Hamilton equations, ${\bar {L_{qi}}}
\cir 0$, have a purely kinematical content: they give the
inversion of the equations $p_i = {\cal P}_i(q, \dot q)$, i.e. ${\dot q}^i = {\bar
g}^i(q,p)$.

By introducing the Poisson brackets\footnote{They satisfy: i) $\{
\bar f, \bar g \} =- \{ \bar g, \bar f \}$; ii) $\{ \bar f, {\bar
g}_1{\bar g}_2 \} = \{ \bar f, {\bar g}_1 \} \, {\bar g}_2 + {\bar
g}_1\, \{ \bar f, {\bar g}_2 \}$ (Leibnitz rule for derivations);
iii) $\{ \{ \bar f, \bar g \} , \bar u \} + \{ \{ \bar g, \bar u
\} , \bar f \} + \{ \{ \bar u, \bar f \} , \bar g \} =0$ (Jacobi
identity).} $ \{ \bar A(q,p),\bar B(q,p) \} = {{\partial \bar A}\over {\partial
 q^i}} \,  {{\partial \bar B}\over {\partial p^i}} -  {{\partial \bar A}\over
 {\partial p^i}}\,  {{\partial \bar B}\over {\partial q^i}}$,
we can re-write the Hamilton equations in the form ${\dot q}^i \cir \{ q^i, \bar H \} = {\bar X}_{\bar H}\, q^i$,
${\dot p}_i \cir \{ p_i, \bar H \} = {\bar X}_{\bar H}\, p_i$,
 where we introduced the evolution Hamiltonian vector
field ${\bar X}_{\bar H} = \{ ., \bar H \}$.

In the regular case every function $f(q, \dot q)$ is {\it
projectable} to phase space: $f(q, \dot q) = \bar f(q,p)$ by using
${\dot q}^i = {\bar g}^i(q,p)$.

Let us remark that there are two intrinsic formulations of the
Hamiltonian description (we consider only the case of an exact
symplectic structure arising when there is a well defined
Lagrangian):

\noindent i) a time-independent one on the {\it symplectic
manifold} $T^*Q$ (the {\it symplectic structure}) based on the
{\it Cartan-Liouville one-form} $\bar \theta = p_i\, dq^i$ and on
the {\it closed symplectic two-form} $\bar \omega =d\bar \theta =
dp_i \wedge dq^i$, $d\bar \omega =0$, where we gave the coordinate expression in Darboux
coordinates adapted to the symplectic structure;
\noindent ii) a time-dependent one on $R \times T^*Q$ (the {\it
contact structure}; $R$ is the time axis; this formulation allows
one to treat also time-dependent Lagrangians, $L(t, q, \dot q)$ )
based on the {\it Poincare'-Cartan one-form} ${\tilde {\bar
\theta}} = \bar L\, d t = \bar \theta - \bar H\, d t$ and on the
{\it closed contact two-form} ${\tilde {\bar \omega}} = d{\tilde
{\bar \theta}} =\bar \omega - d\bar H \wedge d t$.

In the regular case all these descriptions are equivalent.

\subsection{The First Order Velocity Space Formalism.}

When the second order differential equations of motion are in the
normal form (${\ddot q}^i \cir \Lambda^i $), they can be re-written as a set of
first order differential equations on the {\it velocity space} $T
Q$ (the tangent bundle) over Q with coordinates $q^i$, $v^i$
(we shall denote $\tilde f = \tilde f(q,v)$ the functions
on the velocity space and we have $f(q,\dot q) = \tilde f(q,v)
{|}_{v=\dot q}$) with the following position

\beq
 {\dot q}^i \cir v^i,\qquad\qquad
 {\dot v}^i \cir {\tilde \Lambda}^i(q,v).
 \label{2.2}
 \eeq

By introducing $\tilde L(q,v) = L(q,\dot q) {|}_{\dot q=v}$ and
the {\it energy function} $\tilde E = {{\partial \tilde L}\over
{\partial v^i}}\, v^i - \tilde L$, we can define the $T Q$ action
$\tilde S = \int^{t_f}_{t_i} dt\, {\tilde L}_v$ and Lagrangian
${\tilde L}_v = {{\partial \tilde L}\over {\partial v^i}}\,
 {\dot q}^i - \tilde E$, whose stationarity yields the velocity space first order
differential equations of motion (${\tilde A}_{ij}$, ${\tilde
B}^{ij}$, ${\tilde R}_{ij}$ are the $T Q$ expressions of
 $A_{ij}$, $B^{ij}$, $R_{ij}$ respectively)

\bea
 {\tilde L}_{qi} &=& -{\tilde A}_{ij} \Big[ {\dot v}^j + {\tilde
 B}^{jk} \Big( {{\partial \tilde E}\over {\partial q^k}} + {\tilde
 R}_{kh}\, {\tilde B}^{hr}\, {{\partial \tilde E}\over {\partial
 v^r}}\Big) \Big] - {\tilde R}_{ij}\, \Big( {\dot q}^j - v^j \Big)
 \cir 0,\nonumber \\
 &&{}\nonumber \\
 {\tilde L}_{vi} &=& {\tilde A}_{ij}\,  \Big( {\dot q}^j - v^j \Big)
 \cir 0.
 \label{2.3}
 \eea

The normal form of these equations are Eqs.(\ref{2.2}), which
can also be written in the form ${\dot q}^i \cir v^i = \{ q^i, \tilde E \}_L$,
${\dot v}^i \cir - {\tilde B}^{ij}\, \Big(  {{\partial \tilde E}\over
 {\partial q^j}} + {\tilde R}_{jk}\, {\tilde B}^{kh}\,
 {{\partial \tilde E}\over {\partial v^h}}\Big) = {\tilde
 B}^{ij}\, {\tilde \alpha}_j = {\tilde \Lambda}^i = \{ v^i, \tilde
 E \}_L$, where we have introduced the (Lagrangian dependent) $T Q$ Poisson brackets
$ \{ \tilde f, \tilde g \}_L = {\tilde B}^{ij} \Big(
 {{\partial \tilde f}\over {\partial q^i}}
  {{\partial \tilde g}\over {\partial v^j}} -
   {{\partial \tilde f}\over {\partial v^j}}
    {{\partial \tilde g}\over {\partial q^i}} \Big) -
     {{\partial \tilde f}\over {\partial v^i}}\, {\tilde B}^{ih}
     {\tilde R}_{hk}\, {\tilde B}^{kj}\,
      {{\partial \tilde g}\over {\partial v^j}}$,
$\{ q^i, q^j \}_L = 0$, $\{ q^i, v^j \}_L = {\tilde
  B}^{ij}$, $\{ v^i, v^j \}_L = {\tilde B}^{ih}{\tilde
  R}_{hk}{\tilde B}^{kj}$.

Therefore we have the same symplectic structure in $T Q$ and
$T^*Q$. In this context the Legendre transformation is defined as
the {\it fiber derivative} $F L$ of $\tilde L(q,v)$: it is a
linear and fiber preserving mapping from $T Q$ to $T^*Q$ defined
by $F L:\, (q,v) \in T_qQ\, \mapsto\, p_i\, dq^i\, \in T_q^*Q$
($p_i = {{\partial \tilde L}\over {\partial v^i}}$). When $F L$ is
a global diffeomorphism of $T Q$, the Lagrangian $\tilde L(q,v)$
is said to be {\it hyper-regular} and one has a global Hamiltonian
formalism (i.e. $\bar H(q,p)$, the Legendre transform of $\tilde
E(q,v)$, exists globally on $T^*Q$). When $F L$ is only a local
diffeomorphism of $T Q$, $\tilde L(q,v)$ is said to be {\it
regular} and $\bar H(q,p)$ exists only locally. In the regular case $F L$ is a
symplectomorphism which connects the symplectic structures of
$T^*Q$ and $T Q$.

The definition ${\dot q}^i \cir v^i \equiv \tilde \Gamma\, q^i$ and
Eqs.(\ref{2.2}) imply ${\ddot q}^i \cir {{d v^i}\over {d t}}
\cir {\tilde \Lambda}^i = \tilde \Gamma\, v^i$ in the regular
case: this is called the {\it second order differential equation
(SODE) condition} ensuring that $\tilde \Gamma$ is a second order
vector field. See Ref.\cite{23} for the study of the phase space over the
velocity space, i.e. $T^*(T Q)$.

\section{Singular Lagrangians and Hamiltonian constraints for Finite-Dimensional Systems}

Let us consider a finite-dimensional system with a $n$-dimensional
configuration space $Q$ admitting a global coordinate system
$q^i$, $i=1,..,n$ \footnote{Otherwise the following treatment will
only hold locally in a chart of the coordinate atlas of $Q$.}. Let
its dynamics be described by a time-independent singular
Lagrangian $L(q(t), \dot q(t))$ (the extension to the
time-dependent case does not introduce further complications),
namely such that its Hessian matrix is singular: $det\,
{{\partial^2L(q, \dot q)}\over {\partial {\dot q}^i \partial {\dot
q}^j}}\, = 0$.

In this Section we introduce the Hamiltonian formalism for singular
systems and then we shall come back to study the second order
Lagrangian formalism, giving also some information about the
velocity space formalism in Section V after having looked at the notion of DO in Section IV. We shall follow
Refs.\cite{6,23,24,25}.

\subsection{Primary Hamiltonian Constraints and the Hamilton-Dirac
Equations.}

When the Hessian matrix is singular the EL equations (\ref{2.1})
cannot be put in normal form. This means that the accelerations
${\ddot q}^i$ cannot be uniquely determined in terms of $q^i$,
${\dot q}^i$ and that the solutions of the EL equations may depend
on arbitrary functions of time.

Moreover $det\, A_{ij}(q, \dot q) =0$ implies that the canonical
momenta cannot be inverted to get the velocities
${\dot q}^i$ in terms of $q^i$, $p_i$. The $n$ functions $p_i =
{\cal P}_i(q, \dot q)$ are not functionally independent, namely
there are as many identities $\phi_A(q, {\cal P}(q, \dot q)) \equiv 0$, A=1,..,m,
as null eigenvalues of the Hessian matrix. In phase
space ($T^*Q$) these identities become the {\it primary
Hamiltonian constraints} (their functional form  is highly arbitrary)

\beq
 {\bar \phi}_A(q, p) =0,\quad  A=1,..,m,
 \label{3.1}
 \eeq

\noindent which identify the region $\gamma$ of $T^*Q$ allowed to
the configurations of the singular system. Points outside $\gamma$
are not accessible, but we go on to work in $T^*Q$ to utilize its
symplectic structure ($\gamma$ in general has not such a
structure), i.e. its Poisson brackets.

Eq.(\ref{3.1}) is usually written with Dirac's {\it weak
equality sign} $\approx$, i.e.   ${\bar \phi}_A(q, p) \approx 0$, A=1,..,m.
An equation $\bar f(q, p) \approx 0$ means that the
function $\bar f$ vanishes on $\gamma$, but can be different from
zero outside $\gamma$ so that it cannot be put equal to zero
inside the $T^*Q$ Poisson brackets even when they are restricted
to $\gamma$. Instead the {\it strong equality symbol} $\equiv$
(like for {\it identical}) is used for a function $\bar f(q, p)$
vanishing on $\gamma$, $\bar f \approx 0$, and such that also its
differential vanishes on $\gamma$, $d\bar f \approx 0$; such a
function (for instance $\bar f = {\bar \phi}_A^2$) can be put
equal to zero inside Poisson brackets restricted to $\gamma$.

Let us assume that the rank of the Hessian matrix $A_{ij}(q, \dot
q)$ is {\it constant} and equal to $m$ for every value of $(q,
\dot q)$. See Section V and Ref.\cite{24} for what may happen when we relax this
assumption.

Let us also assume that the singular Lagrangian is such that the
region $\gamma$ defined by the primary constraints is a $(2n -
m)$-dimensional {\it sub-manifold} of $T^*Q$ ($\gamma$ is the {\it
primary constraint sub-manifold}). While $p_k \approx 0$ is an acceptable constraint,
neither $p^2_k \approx 0$ nor $\sqrt{p_k} \approx 0$ are
acceptable. When the rank of the Hessian matrix is not constant (see Section V),
constraints of the type $p^2_k \approx 0$ may appear. Constraints of the type $(p_k)^2
+ (q^h)^2 \approx 0$ must be put in the form $p_k \approx 0$ and $q^h \approx 0$.

While in the regular case the invertibility of the equations $p_i
= {\cal P}(q, \dot q)$ to ${\dot q}^i = {\bar g}^i(q, p)$ implies
that all the velocities ${\dot q}^i$ are {\it projectable} to
$T^*Q$, now there will be $m$ independent (but with a not uniquely
determined functional form) functions of the velocities $g^A(q,
\dot q)$ (named {\it non-projectable velocity functions}) not
projectable to $T^*Q$.

\bigskip

Let us also assume that the singular Lagrangian admits a well
defined Legendre transformation. Then, if we introduce the
function $H_c(q, \dot q) = p_i\, {\dot q}^i - L(q, \dot q) = {\cal P}_i(q,
 \dot q)\, {\dot q}^i - L(q, \dot q)$,
 we get $\delta H_c = \delta p_i\, {\dot q}^i + p_i\,
 \delta {\dot q}^i - {{\partial L}\over {\partial q^i}}\, \delta
 q^i - {{\partial L}\over {\partial {\dot q}^i}}\, \delta {\dot
 q}^i = {\dot q}^i\, \delta p_i - {{\partial L}\over {\partial
 q^i}}\, \delta q^i = \delta {\bar H}_c$ as in the regular case.
 This means that $H_c(q, \dot q)$ is projectable to a well defined
 {\it canonical Hamiltonian} also in the singular case

 \beq
 H_c(q, \dot q) = {\bar H}_c(q, p),\quad \Big( {{\partial {\bar
 H}_c}\over {\partial q^i}} + {{\partial L}\over {\partial
 q^i}}\Big)     \, \delta q^i + \Big( {{\partial {\bar H}_c}\over
 {\partial p_i}} - {\dot q}^i\Big)\, \delta p_i =0.
 \label{3.2}
 \eeq

But in the singular case Eqs.(\ref{3.2}) are meaningful only if
$(\delta q^i, \delta p_i)$ is a vector tangent to the primary
constraint sub-manifold $\gamma$, so that one gets (see Ref.\cite{6} for a demonstration)

\bea
 {\dot q}^i &=& {{\partial {\bar H}_c}\over {\partial p_i}} +
 u^A\, {{\partial {\bar \phi}_A}\over {\partial p_i}} = \{ q^i,
 {\bar H}_c \} + u^A\, \{ q^i, {\bar \phi}_A \} ,\nonumber \\
 &&{}\nonumber \\
 {\dot p}_i &=& {{\partial L}\over {\partial q^i}} {|}_{\dot q}\,
 = - {{\partial {\bar H}_c}\over {\partial q^i}} - u^A\,
 {{\partial {\bar \phi}_A}\over {\partial q^i}} = \{ p_i, {\bar
 H}_c \} + u^A\, \{ p_i, {\bar \phi}_A \},
 \label{3.3}
 \eea

\noindent where the EL equations have been used in the second
line. Since the velocities are not projectable to $T^*Q$ in the singular
case, the functions $u^A$ in the first line of Eqs.(\ref{3.3})
cannot be functions on $T^*Q$ but must depend also on the
velocities: $u^A = u^A(q, p, \dot q) = u^A(q, {\cal P}(q, \dot q),
\dot q) = v^A(q, \dot q)$. These multipliers identify a {\it
canonical functional form} $g^A_{(u)} = v^A$ of the
non-projectable velocity functions $g^A(q, \dot q)$.

However, since Eqs.(\ref{3.3}) are the Hamilton equations for
the singular case (the so called {\it Hamilton-Dirac equations}),
their right side cannot depend explicitly on the velocities.
Therefore a consistent Hamiltonian formalism is obtained by
replacing the functions $u^A = v^A(q, \dot q)$ with arbitrary
multipliers $\lambda^A(t)$ (the so called {\it Dirac
multipliers}), by introducing the {\it Dirac Hamiltonian}

\beq
 {\bar H}_D(q, p, \lambda ) = {\bar H}_c(q, p) + \lambda^A(t)\, {\bar
 \phi}_A(q, p),
 \label{3.4}
 \eeq

\noindent and by introducing the $T^*Q$ action
$\bar S = \int^{t_f}_{t_i}\, d t\, ( p_i\, {\dot q}^i - {\bar H}_c
 - \lambda^A(t)\, {\bar \phi}_A)$.

In it $q^i$, $p_i$ and $\lambda^A$ are considered as independent
variables. The stationarity, $\delta \bar S =0$, under variations
$\delta q^i$, $\delta p_i$, $\delta \lambda^A$ with the only
restriction $\delta q^i(t_f) = \delta q^i (t_i) =0$ yields the
Hamilton-Dirac equations supplemented by the definition of the
primary constraint sub-manifold

 \bea
  {\dot q}^i &\cir& \{ q^i, {\bar H}_D(q, p, \lambda ) \},\quad {\it or}\quad
  {\bar L}^i_{Dq}\, = {\dot q}^i - \{ q^i, {\bar H}_D \} \cir 0,\nonumber \\
  &&{}\nonumber \\
  {\dot p}_i &\cir& \{ p_i, {\bar H}_D(q, p, \lambda ) \},\quad {\it or}\quad
  {\bar L}_{Dpi}\, = {\dot p}_i - \{ p_i, {\bar H}_D \} \cir 0,\nonumber \\
  &&{}\nonumber \\
  {\bar \phi}_A(q, p) &\cir& 0.
  \label{3.5}
  \eea

The kinematical equations ${\dot q}^i \cir \{ q^i, {\bar H}_D \}$,
${\bar \phi}_A \cir 0$, determine the canonical momenta and the
Dirac multipliers in terms of the coordinates and momenta (if
suitable regularity conditions hold), namely we get: i) $p_i =
{\cal P}_i(q, \dot q)$; ii) the canonical functional form
$g^A_{(\lambda )}$ of the non-projectable velocity functions
$g^A(q, \dot q)$ associated with the chosen functional form of the
primary constraints as the $(q, \dot q)$ space expression of the
Dirac multipliers $g^A_{(\lambda )}(q, \dot q) \cir \lambda^A(t)$.
Then we can make the {\it inverse Legendre transformation} and
recover the original singular Lagrangian: $p_i\, {\dot q}^i -
{\bar H}_D \cir {\cal P}_i(q, \dot q)\, {\dot q}^i - {\bar H}_c(q,
{\cal P}(q, \dot q)) = L(q, \dot q)$.

\subsection{Dirac's Algorithm for the Determination of the Final
Constraint Sub-manifold.}

The inspection of the functional form of the canonical momenta
$p_i = {\cal P}_i(q, \dot q)$ identifies the primary constraint
sub-manifold $\gamma \subset T^*Q$. The Hamiltonian formalism will
produce a consistent treatment of singular systems only if
$\gamma$ does not change with time, namely if the
primary constraints ${\bar \phi}_A(q, p) \approx 0$ are constant
of motion with respect to the evolution generated by the Dirac
Hamiltonian

\bea
 {{d {\bar \phi}_A(q, p)}\over {d t}} &\cir& \{ {\bar \phi}_A(q,
 p), {\bar H}_D \} =\nonumber \\
 &=& \{ {\bar \phi}_A(q, p), {\bar H}_c(q, p) \} + \lambda^B(t)\,
 \{ {\bar \phi}_A(q, p), {\bar \phi}_B(q, p) \} \approx 0\quad
 {\it on\,\, \gamma},\,\, A=1,..,m.\nonumber \\
 &&{}
 \label{3.6}
 \eea

Some of these equations may be void ($0=0$). The non-void ones,
restricted to $\gamma$, have to be separated in two disjoint
groups:

\noindent i) a set of $m_1 \leq m$ equations independent from the
Dirac multipliers

 \beq
  {\bar \chi}_{a_1}^{(1)}(q, p)\, \approx 0\quad {\it on\,
  \gamma},\quad a_1=1,..,m_1;
  \label{3.7}
  \eeq

\noindent ii) a set of $h_1$ equations ($h_1 \leq m$, $h_1 + m_1
\leq m$) for the Dirac multipliers

\beq
 {\bar f}_{{\tilde A}_1 B}(q, p)\, \lambda^B(t) + {\bar
 g}_{{\tilde A}_1}(q, p) \approx 0\quad {\it on\, \gamma},\quad
 {\tilde A}_1=1,..,h_1,
 \label{3.8}
 \eeq

\noindent with ${\bar f}_{{\tilde A}_1} = {\bar h}^A_{{\tilde
A}_1}\, \{ {\bar \phi}_A, {\bar \phi}_B \}$, ${\bar g}_{{\tilde
A}_1} = {\bar h}^A_{{\tilde A}_1}\, \{ {\bar \phi}_A, {\bar H}_c
\}$ for some functions ${\bar h}^A_{{\tilde A}_1}$.

Let us remark that without certain regularity conditions on the
singular Lagrangian this separation cannot be done in a unique
way: i) we can get different separations in different regions of
$\gamma$; ii) also in the same point of $\gamma$ we can have
alternative inequivalent separations. A regularity
condition which eliminates most (if not all) of these
possibilities is that the anti-symmetric matrix ${\bar M}_{AB}(q,
p) = \Big( \{ {\bar \phi}_A(q, p),\, {\bar \phi}_B(q, p) \} \Big)$
of the Poisson brackets of the primary constraints has {\it
constant rank on $\gamma$}.

Let us assume that there is a unique separation given by Eqs.
(\ref{3.7}), (\ref{3.8}). Eqs. (\ref{3.7}) are called
{\it secondary constraints} and define a {\it secondary constraint
sub-manifold} $\gamma_1$ of $\gamma$ to which the description of
the singular system has to be restricted for consistency.
Differently from the primary constraints, the secondary
constraints are defined by using the equations of motion.
Eqs.(\ref{3.8}) show that on the constraint sub-manifold
$\gamma_1 \subset \gamma \subset T^*Q$ there may be less
arbitrariness than on $\gamma$, because $h_1 \leq m$ Dirac
multipliers $\lambda^A(t)$ are determined by these equations and
an equal number of velocity functions $g^A_{(\lambda )}(q, \dot
q)$, not projectable onto $\gamma$, become projectable onto
$\gamma_1 \subset \gamma$.

If ${\bar U}^A(q, p)$ is a particular solution of the
in-homogeneous Eqs.(\ref{3.8}) and ${\bar V}^A_{A_1}(q, p)$,
$A_1=1,..,k_1 = m-h_1$, are independent solutions of the
homogeneous equations ${\bar f}_{{\tilde A}_1B}\, \lambda^B(t) =
{\bar h}^A_{{\tilde A}_1}\, \{ {\bar \phi}_A, {\bar \phi}_B \}\,
\lambda^B(t) =0$ (namely $\{ {\bar \phi}_A, {\bar \phi}_B \}\,
{\bar V}^B_{A_1} \equiv 0$), then the general solution of
Eqs.(\ref{3.8}) is

\beq
 \lambda^A(t)\, \approx \, {\bar U}^A(q, p) + \lambda^{(1)\,
 A_1}(t)\, {\bar V}^A_{A_1}(q, p)\quad {\it on\, \gamma_1},\quad
 A_1=1,..,k_1=m-h_1,
 \label{3.9}
 \eeq

\noindent with the $\lambda^{(1)A_1}(t)$'s being new $k_1=m-h_1$
arbitrary Dirac multipliers.

The Dirac Hamiltonian on $\gamma_1 \subset \gamma$ is

 \bea
 {\bar H}_D(q, p, \lambda ) &=& {\bar H}_c^{(1)}(q, p) +
 \lambda^{(1)A_1}(t)\, {\bar \phi}^{(1)}_{A_1}(q, p)\quad {\it
 on\, \gamma_1},\nonumber \\
 &&{}\nonumber \\
 {\bar \phi}^{(1)}_{A_1}(q, p) &=& {\bar V}^A_{A_1}(q, p)\, {\bar
 \phi}_A(q, p),\quad A_1=1,..,k_1 =m - h_1,\nonumber \\
 {\bar H}^{(1)}_c(q, p) &=& {\bar H}_c(q, p) + {\bar U}^A(q, p)\,
 {\bar \phi}_A(q, p) \approx {\bar H}_c(q, p) \quad {\it on\,
 \gamma_1}.
 \label{3.10}
 \eea

The remaining Dirac multipliers $\lambda^{(1)A_1}(t)$ are now
multiplied by the linear combinations ${\bar \phi}^{(1)}_{A_1}(q,
p)$ of the original primary constraints.

When there are secondary constraints ${\bar \chi}^{(1)}_{a_1}(q,
p) \approx 0$, for consistency we must ask that the secondary
constraint sub-manifold $\gamma_1 \subset \gamma$ does not change
with time: the secondary constraints must be constants of motion
on $\gamma_1$ with respect to the Dirac Hamiltonian

\bea
 {{d {\bar \chi}^{(1)}_{a_1}(q, p)}\over {d t}} &\cir& \{ {\bar \chi}^{(1)}
 _{a_1}(q, p), {\bar H}_D(q, p, \lambda ) \} \approx \nonumber \\
 &\approx& \{ {\bar \chi}^{(1)}_{a_1}(q, p), {\bar H}^{(1)}_c(q,
 p) \} + \lambda^{(1)\, A_1}(t)\, \{ {\bar \chi}^{(1)}_{a_1}(q,
p), {\bar \phi}^{(1)}_{A_1}(q, p) \} \approx 0\quad {\it on\,\,
\gamma_1},\nonumber \\
 &&A_1=1,.., k_1 = m - h_1,\quad a_1 = 1,..,m_1.
  \label{3.11}
   \eea

By assuming the regularity condition that the rank of the matrix
$\Big( \{ {\bar \chi}^{(1)}_{a_1}, {\bar \phi}^{(1)}_{A_1}
\}\Big)$ is {\it constant on $\gamma_1$}, the non-void
Eqs.(\ref{3.11}) may be separated in the two disjoint sets

\beq
 {\it i)}\quad {\bar \chi}^{(2)}_{a_2}(q, p)\, \approx 0\quad {\it
 on\,\, \gamma_1},\quad a_2=1,..,m_2 \leq m_1;
 \label{3.12}
 \eeq

 \beq
 {\it ii)}\quad {\bar f}_{{\tilde A}_2B_1}(q, p)\,
 \lambda^{(1)B_1}(t) + {\bar g}_{{\tilde A}_2}(q, p) \approx
 0\quad {\it on\,\, \gamma_1},\quad {\tilde A}_2=1,..,h_2 \leq
 k_1.
 \label{3.13}
 \eeq

Eqs.(\ref{3.12}) are the {\it tertiary constraints} and define a
new constraint sub-manifold $\gamma_2 \subset \gamma_1$. With the
same procedure delineated above we arrive at the conclusion that
only on $\gamma_2$ can there be a consistent dynamics for the
singular system with (in general) a reduction of the number of
independent Dirac multipliers, which are replaced by the new ones
$\lambda^{(2)A_2}(t)$ due to Eqs.(\ref{3.13}). The dynamics is
described in terms of the following quantities

\bea
 \lambda^{(1)A_1}(t) &\approx& {\bar U}^{(1)A_1}(q, p) +
 \lambda^{(2)A_2}(t)\, {\bar V}^{(1)}{}^{A_1}_{A_2}(q, p)\quad
 {\it on\,\, \gamma_2},\quad A_2=1,..,k_2=m-h_1-h_2,\nonumber \\
 &&{}\nonumber \\
 {\bar H}_D(q, p, \lambda ) &=& {\bar H}^{(2)}_c(q, p) +
 \lambda^{(2)A_2}(t)\, {\bar \phi}^{(2)}_{A_2}(q, p),\quad
 {\it on\,\, \gamma_2},\nonumber \\
 &&{}\nonumber \\
 {\bar \phi}^{(2)}_{A_2}(q, p) &=& {\bar V}^{(1)}{}^{A_1}_{A_2}(q,
 p)\, {\bar \phi}^{(1)}_{A_1}(q, p) = {\bar
 V}^{(1)}{}^{A_1}_{A_2}(q, p)\, {\bar V}^A_{A_1}(q, p)\,
 {\bar \phi}_A(q, p),\nonumber \\
 {\bar H}_c^{(2)}(q, p) &=& {\bar H}^{(1)}_c(q, p) + {\bar
 U}^{(1)A_1}(q, p) {\bar \phi}^{(1)}_{A_1}(q, p) =\nonumber \\
 &=& {\bar H}_c(q, p) + \Big( {\bar U}^A(q, p) + {\bar
 U}^{(1)A_1}(q, p)\, {\bar V}^A_{A_1}(q, p)\Big)\, {\bar
 \phi}_A(q, p).
 \label{3.14}
 \eea

This procedure is iterated till a final stage $(f)$ in which the
{\it final constraint sub-manifold} $\bar \gamma = \gamma_f
 \subset ... \subset \gamma_1 \subset \gamma \subset T^*Q$ is
 determined by {\it $(f + 1)$-ary constraints}

\beq
 {\bar \chi}^{(f)}_{a_f}(q, p)\, \approx 0\quad {\it on\,
 \gamma_{f-1}}, \quad a_f = 1,..,m_f \leq m_{f-1} \leq ... \leq m.
 \label{3.15}
 \eeq

On $\bar \gamma = \gamma_f$ we have

\bea
 {\bar H}_D(q, p, \lambda ) &=& {\bar H}^{(f)}_c(q, p) +
 \lambda^{(f)A_f}(t)\, {\bar \phi}^{(f)}_{A_f}(q, p),\quad
 {\it on\,\, \gamma_f},\nonumber \\
 &&{}\nonumber \\
 {\bar \phi}^{(f)}_{A_f}(q, p) &=&  {\bar
 V}^{(f-1)}{}^{A_{f-1}}_{A_f}(q, p)\, ...\, {\bar V}^A_{A_1}(q, p)\,
 {\bar \phi}_A(q, p),\quad A_f=1,..,k_f=m-h_1-..-h_f,\nonumber \\
 {\bar H}_c^{(f)}(q, p) &=& {\bar H}_c(q, p) + \Big( {\bar U}^A(q, p) + {\bar
 U}^{(1)A_1}(q, p)\, {\bar V}^A_{A_1}(q, p) + ... +\nonumber \\
 &+& {\bar U}^{(f-1)A_{f-1}}(q, p)\, {\bar V}^{(f-2)}{}^{A_{f-2}}_{A_{f-1}}(q, p)\,
 ...\, {\bar V}^A_{A_1}(q, p)\Big)\, {\bar \phi}_A(q, p) \approx
 {\bar H}_c(q, p),\nonumber \\
 &&{}\nonumber \\
 {{d {\bar \chi}^{(f)}_{a_f}(q, p)}\over {d t}} &\cir& \{
 {\bar \chi}^{(f)}_{a_f}(q, p), {\bar H}_D(q, p, \lambda ) \}\quad
 {\it identically\,\, satisfied},
 \label{3.16}
 \eea

\noindent with only $k_f = m - h_1 - .. - h_f$ independent final
Dirac multipliers $\lambda^{(f)A_f}(t)$. Only an equal number of
velocity functions $g^A_{(\lambda )}(q, \dot q)$ cannot be
projected onto $\bar \gamma= \gamma_f \subset T^*Q$.

The solutions of the Hamilton-Dirac equations on $\bar \gamma$
will depend {\it on the $k_f$ arbitrary functions of time}
$\lambda^{(f)A_f}(t)$, which describe the {\it non-deterministic}
aspects of the time evolution of the singular system.

\subsection{First and Second Class Constraints.}

The final constraint manifold $\bar \gamma = \gamma_f \subset ..
\subset \gamma \subset T^*Q$ is determined by the full set of
primary, secondary,.. constraints ${\bar \phi}_A(q, p) \approx 0$
($A=1,..,m$), ${\bar \chi}^{(1)}_{a_1}(q, p) \approx 0$ ($a_1
=1,..,m_1$), ..., ${\bar \chi}^{(f)}_{a_f}(q, p) \approx 0$
($a_f=1,..,m_f$). Let us denote all the $M = m + m_1 + .. +  m_f$
constraints with the collective notation

\beq
 {\bar \zeta}_{\cal A}(q, p) \approx 0,\quad {\cal A} = 1,.., M,
 \label{3.17}
 \eeq

\noindent because the property of being a primary, secondary ..
constraint is not important.

Let us also denote with $\lambda^{\bar A}(t)$ and ${\bar
\phi}_{\bar A}(q, p)$ with $\bar A =1,.., \bar k = m - h_1 -..-
h_f$ the final arbitrary Dirac multipliers and the associated
linear combinations of primary constraints respectively, so that
the Dirac Hamiltonian on $\bar \gamma$ is

 \bea
  {\bar H}_D(q, p, \lambda ) &=& {\bar H}^{(F)}_c(q, p) +
 \lambda^{\bar A}(t)\, {\bar \phi}_{\bar A}(q, p),\quad
 {\it on\,\, \bar \gamma},\nonumber \\
 &&{}\nonumber \\
 {\bar \phi}_{\bar A}(q, p) &=&  {\bar
 V}^{(f-1)}{}^{A_{f-1}}_{\bar A}(q, p)\, ...\, {\bar V}^A_{A_1}(q, p)\,
 {\bar \phi}_A(q, p),\quad \bar A=1,..,\bar k,\nonumber \\
 {\bar H}_c^{(F)}(q, p) &=& {\bar H}_c(q, p) + {\bar U}^{(F) A}(q,
 p)\, {\bar \phi}_A(q, p)   \approx {\bar H}_c(q, p),\nonumber \\
 &&{}\nonumber \\
 {\bar U}^{(F) A}(q, p) &=&  {\bar U}^A(q, p) + {\bar
 U}^{(1)A_1}(q, p)\, {\bar V}^A_{A_1}(q, p) + ... +\nonumber \\
 &+& {\bar U}^{(f-1)A_{f-1}}(q, p)\, {\bar V}^{(f-2)}{}^{A_{f-2}}_{A_{f-1}}(q, p)\,
 ...\, {\bar V}^A_{A_1}(q, p).
 \label{3.18}
 \eea

As a result of the previous construction all the constraints are
preserved in time on $\bar \gamma$,

\beq
 {{d {\bar \zeta}_{\cal A}}\over {d t}} \cir \{ {\bar \zeta}_{\cal
A}, {\bar H}_D \} \approx 0
 \Rightarrow\quad \{ {\bar \zeta}_{\cal A}, {\bar H}_c^{(F)} \}
 \approx 0,\quad \{ {\bar \zeta}_{\cal A}, {\bar \phi}_{\bar A} \}
 \approx 0.
 \label{3.19}
 \eeq

Let us call a {\it first class function} a function $\bar f(q, p)$
on $T^*Q$ whose Poisson brackets with every constraint is weakly
zero

\beq
 \{ \bar f(q, p), {\bar \zeta}_{\cal A}(q, p) \} = {\bar F}^{\cal
 B}_{\cal A}(q, p)\, {\bar \zeta}_{\cal B}(q, p) \approx 0\quad {\it
 on\,\, \bar \gamma}.
 \label{3.20}
 \eeq

If two functions $\bar f$, $\bar g$ are first class, also their
Poisson bracket is a first class function due to the Jacobi
identity: $\{\, \{ \bar f, \bar g \} , {\bar \zeta}_{\cal A} \} =
\{ \bar f, \{ \bar g, {\bar \zeta}_{\cal A} \}\, \} - \{ \bar g,
\{ \bar f, {\bar \zeta}_{\cal A} \}\, \} = \{ \bar f, {\bar
G}^{\cal B}_{\cal A}\, {\bar \zeta}_{\cal B} \} - \{ \bar g, {\bar
F}^{\cal B}_{\cal A}\, {\bar \zeta}_{\cal B} \} = {\bar K}^{\cal
B}_{\cal A}\, {\bar \zeta}_{\cal B} \approx 0$.

All the functions which are not  first class are named {\it second
class functions}.

Eqs.(\ref{3.19}) show that both the final canonical Hamiltonian
${\bar H}^{(F)}_c(q, p)$ and the final combinations ${\bar
\phi}_{\bar A}(q, p)$, $\bar A=1,..,\bar k$, of the primary
constraints are first class functions.

It is of fundamental importance in constraint theory to separate
the constraints ${\bar \zeta}_{\cal A} = \Big( {\bar \phi}_A,\,
{\bar \chi}^{(1)}_{a_1},\, .., \,$ $ {\bar
\chi}^{(f)}_{a_f}\Big)$ in two groups: i) the {\it first class
constraints} ${\bar \Phi}_{(1){\cal A}_1} = {\bar k}^{\cal
A}_{{\cal A}_1}\, {\bar \zeta}_{\cal A} \approx 0$, ${\cal
A}_1=1,..,r_1$, $\{ {\bar \Phi}_{(1){\cal A}_1}, {\bar
\zeta}_{\cal A} \} \approx 0$; ii) the {\it second class
constraints} ${\bar \Phi}_{(2){\cal A}_2} \approx 0$, ${\cal
A}_2=1,.., r_2$ ($r_1+r_2=M$) with $\{ {\bar \Phi}_{(2){\cal
A}_2}, {\bar \zeta}_{\cal A} \} \not= 0$ for some ${\cal A}$.
Evidently we have $\{ {\bar \Phi}_{(1){\cal A}_1}, {\bar
\Phi}_{(1){\cal B}_1} \} \approx 0$, $\{ {\bar \Phi}_{(1){\cal
A}_1}, {\bar \Phi}_{(2){\cal A}_2} \} \approx 0$, $det\, \Big( \{
{\bar \Phi}_{(2){\cal A}_2}, {\bar \Phi}_{(2){\cal B}_2}   \}
\Big) \not= 0$.

The ${\bar \phi}_{\bar A}$'s constitute a {\it complete set of
first class primary constraints}.

If a set of first class constraints has the form ${\bar
\Phi}_{(1)a} = p_a - {\bar K}_a(q^a, q^r, p_r) \approx 0$ with $r
\not= a$ (i.e. they are solved in a subset of the momenta), then
$\{ {\bar \Phi}_{(1)a}, {\bar \Phi}_{(1)b} \} = {{\partial {\bar
K}_a}\over {\partial q^b}} - {{\partial {\bar K}_b}\over {\partial
q^a}} + \{ {\bar K}_a, {\bar K}_b \} \equiv 0$.

Geometrically the Hamiltonian vector fields ${\bar X}_{(1){\cal
A}_1} = \{ ., {\bar \Phi}_{(1){\cal A}_1} \}$ and  ${\bar
X}_{(2){\cal A}_2} = \{ ., {\bar \Phi}_{(2){\cal A}_2} \}$ are
{\it tangent and skew} respectively to the constraint sub-manifold
$\bar \gamma$ (see Ref.\cite{6}; in Ref.\cite{26} there is a study of the conditions
for putting all the first-class constraints in this Abelianized form as also discussed in Subsection IVC).

\subsection{Chains of Constraints: Diagonalization of the Dirac Algorithm.}

We quote three theorems \cite{14,27,28} on equivalent sets of
constraints  ${\bar \zeta}_{\cal A}(q, p) \approx 0$, ${\tilde
{\bar \zeta}}_{\cal A}(q, p) \approx 0$ both defining $\bar
\gamma$, valid when suitable regularity conditions hold, without
reproducing the long not illuminating demonstrations based on
inductive procedures. These theorems allow to perform a {\it
diagonalization} of the Dirac algorithm and to separate the
constraints in {\it chains} (one for each primary constraint,
namely for each null eigenvalue of the Hessian matrix), such that
the time constancy of a constraint in the chain implies the next
constraint in the chain. The time constancy of the last constraint
in the chain either is automatically satisfied or determines the
Dirac multiplier associated to the chain. A {\it $0$-chain} has
only the primary constraint, a {\it $1$-chain} has the primary and
a secondary, and so on.

The first theorem shows the existence of diagonalized chains.

\noindent {\it Theorem 1} \cite{14} - By taking suitable
combinations ${\bar \phi}_{\bar A} = {\bar
V}^{(f-1)}{}^{A_{f-1}}_{\bar A}\, ...\, {\bar V}^A_{A_1}\, {\bar
\phi}_A$ ($\bar A=1,..,\bar k=m-h_1-..-h_f$) and ${\bar
\phi}_{A^{'}} = {\bar u}^A_{A^{'}}\, {\bar \phi}_A$
($A^{'}=1,..,m-\bar k$) of the primary constraints ${\bar \phi}_A$
($A=1,..,m$) defining $\gamma$, then suitable combinations of the
secondary ${\bar \chi}^{(1)}_{a_1}$ and primary ${\bar \phi}_A$
constraints  defining $\gamma_1$ and so on, the final pattern of
the chains of constraints can be put in the following form:

\noindent i) chains of constraints starting from primary
constraints whose Dirac multiplier $\lambda^A(t)$ is determined by
the Dirac algorithm on $\bar \gamma$. We use the following
notation: ${\bar \phi}_{(h)A^{'}_h} \approx 0$ is the primary
constraint of a $h$-chain of $h+1$ constraints ($A^{'}_h$ labels
the various $h$-chains), ${\bar \phi}^{(1)}_{(h)A^{'}_h}\,
\approx\, {{d {\bar \phi}_{(h)A^{'}_h}}\over {d t}}\, \approx \{
{\bar \phi}_{(h)A^{'}_h}, {\bar H}^{(1)}_c \}\, \approx 0$ is the
secondary,  ${\bar \phi}^{(2)}_{(h)A^{'}_h}\, \approx\, {{d {\bar
\phi}^{(1)}_{(h)A^{'}_h}}\over {d t}}\, \approx \{ {\bar
\phi}^{(1)}_{(h)A^{'}_h}, {\bar H}^{(2)}_c \}\, \approx 0$ is the
tertiary and so on till  ${\bar \phi}^{(f_h)}_{(h)A^{'}_h}\,
\approx\, {{d {\bar \phi}^{(f_h-1)}_{(h)A^{'}_h}}\over {d t}}\,
\approx \{ {\bar \phi}^{(f_h-1)}_{(h)A^{'}_h}, {\bar H}^{(f_h)}_c
\}\, \approx 0$; then ${{d {\bar \phi}^{(f_h)}_{(h)A^{'}_h}}\over
{d t}}\, \approx 0$ determines the Dirac multiplier. Here ${\bar
H}_D$, ${\bar H}_c$, ${\bar H}^{(1)}_c$,.. are the quantities
already introduced in the previous Section.

\bea
 \begin{array}{llll|l}
 {\it 0-chains} & {\it 1-chains} &  {} & {\it f-chains} & {}\\
 {\bar \phi}_{(o)A^{'}_o}\,\,\, (A^{'}_o=1,..,k_o^{'}) &
  {\bar \phi}_{(1)A^{'}_1}\,\,\, (A^{'}_1=1,..,k_1^{'}) & ... &
   {\bar \phi}_{(f)A^{'}_f}\,\,\, (A^{'}_f=1,..,k_f^{'}) &
   {\it primary}\\
  \lambda^{(o)A^{'}_o}\,\, {\it determined} & {\bar
  \phi}^{(1)}_{(1)A^{'}_1} & ... & {\bar
  \phi}^{(1)}_{(f)A^{'}_f} & {\it secondary}\\
  {} & \lambda^{(1)A^{'}_1}\,\, {\it determined} & ... &
  {\bar \phi}^{(2)}_{(f)A^{'}_f} & {\it tertiary}\\
  ... & ... &  ... & ... & {}\\
  {} & {} & ... & {\bar \phi}^{(f)}_{(f)A^{'}_f} & {\it
  (f+1)-ary}\\
  {} & {} & ... & \lambda^{(f)A^{'}_f}\,\, {\it determined}
  \end{array} ;
 \label{3.21}
 \eea

\noindent ii) chains of constraints starting from the primary
constraints ${\bar \phi}_{\bar A}$ with associated arbitrary Dirac
multipliers on $\bar \gamma$. The same notation as in i) is used;
but now  ${{d {\bar \phi}^{(f_h)}_{(h){\bar A}_h}}\over {d t}}\,
\approx 0$ is identically satisfied without determining the Dirac
multiplier.

\bea
 \begin{array}{llll|l}
 {\it 0-chains} & {\it 1-chains} &  {} & {\it f-chains} & {}\\
 {\bar \phi}_{(o){\bar A}_o}\,\,\, ({\bar A}_o=1,..,k_o^{'}) &
  {\bar \phi}_{(1){\bar A}_1}\,\,\, ({\bar A}_1=1,..,k_1^{'}) & ... &
   {\bar \phi}_{(f){\bar A}_f}\,\,\, ({\bar A}_f=1,..,k_f^{'}) &
   {\it primary}\\
  {} & {\bar \phi}^{(1)}_{(1){\bar A}_1} & ... & {\bar
  \phi}^{(1)}_{(f){\bar A}_f} & {\it secondary}\\
  {} & {} & ... &
  {\bar \phi}^{(2)}_{(f){\bar A}_f} & {\it tertiary}\\
  ... & ... &  ... & ... & {}\\
  {} & {} & ... & {\bar \phi}^{(f)}_{(f){\bar A}_f} & {\it
  (f+1)-ary}
  \end{array} .
 \label{3.22}
 \eea

The second theorem shows that the diagonalized chains of Theorem 1
can be redefined so that each  chain has all the constraints
either first or second class.

\noindent {\it Theorem 2} \cite{27} - By leaving the primary
constraints in the form of Theorem 1, we can take linear
combinations of all the other constraints so to obtain the
following pattern:

\noindent i) chains of {\it second class} constraints ${\bar
\chi}^{(h)}_{(k)A^{'}_k}$ with the associated Dirac multiplier
determined \hfill\break ($det\, \Big( \{ {\bar
\chi}^{(h)}_{(k)A^{'}_k}, {\bar \chi}^{(h_1)}_{(k_1)A^{'}_{k_1}}
\} \Big) \not= 0$).

\bea
 \begin{array}{llll|l}
 {\it 0-chains} & {\it 1-chains} &  {} & {\it f-chains} & {}\\
 {\bar \phi}_{(o)A^{'}_o} = {\bar \chi}^{(o)}_{(o)A^{'}_o} &
  {\bar \phi}_{(1)A^{'}_1} = {\bar \chi}^{(o)}_{(1)A^{'}_1} & ... &
   {\bar \phi}_{(f)A^{'}_f}= {\bar \chi}^{(o)}_{(f)A^{'}_f} &
   {\it primary}\\
  \lambda^{(o)A^{'}_o}\,\, {\it determined} & {\bar
  \chi}^{(1)}_{(1)A^{'}_1} & ... & {\bar
  \chi}^{(1)}_{(f)A^{'}_f} & {\it secondary}\\
  {} & \lambda^{(1)A^{'}_1}\,\, {\it determined} & ... &
  {\bar \chi}^{(2)}_{(f)A^{'}_f} & {\it tertiary}\\
  ... & ... &  ... & ... & {}\\
  {} & {} & ... & {\bar \chi}^{(f)}_{(f)A^{'}_f} & {\it
  (f+1)-ary}\\
  {} & {} & ... & \lambda^{(f)A^{'}_f}\,\, {\it determined}
  \end{array} ;
 \label{3.23}
 \eea

\noindent ii) chains of {\it first class} constraints ${\bar
\chi}^{(h)}_{(k){\bar A}_k}$ with arbitrary Dirac multipliers and
with the property ${\bar \chi}^{(h+1)}_{(k){\bar A}_k} = \{ {\bar
\chi}^{(h)}_{(k){\bar A}_k}, {\bar H}^{(F)}_c \}$.

\bea
 \begin{array}{llll|l}
 {\it 0-chains} & {\it 1-chains} &  {} & {\it f-chains} & {}\\
 {\bar \phi}_{(o){\bar A}_o} = {\bar \chi}^{(o)}_{(o){\bar A}_o} &
  {\bar \phi}_{(1){\bar A}_1} = {\bar \chi}^{(o)}_{(1){\bar A}_1} & ... &
   {\bar \phi}_{(f){\bar A}_f} = {\bar \chi}^{(o)}_{(f){\bar A}_f} &
   {\it primary}\\
  {} & {\bar \chi}^{(1)}_{(1){\bar A}_1} & ... & {\bar
  \chi}^{(1)}_{(f){\bar A}_f} & {\it secondary}\\
  {} & {} & ... &
  {\bar \chi}^{(2)}_{(f){\bar A}_f} & {\it tertiary}\\
  ... & ... &  ... & ... & {}\\
  {} & {} & ... & {\bar \chi}^{(f)}_{(f){\bar A}_f} & {\it
  (f+1)-ary}
  \end{array} .
 \label{3.24}
 \eea

The third theorem gives a simple {\it canonical form for the
chains of second class constraints}.

\noindent {\it Theorem 3} \cite{28} - By adding suitable terms
quadratic in the second class constraints to ${\bar H}^{(F)}_c$
(this is irrelevant on $\bar \gamma$) to get a new ${\bar
H}_c^{(F){'}}$ and by making an appropriate linear orthogonal
transformation on the set of primary second class constraints $
{\bar \phi}_{(k)A^{'}_k}= {\bar \chi}^{(o)}_{(k)A^{'}_k}$ to
obtain the new primary second class constraints $ {\bar
\chi}^{(o){'}}_{(k)A^{'}_k}$, we get new forms $ {\bar
\chi}^{(h){'}}_{(k)A^{'}_k}$ of the non-primary second class
constraints such  that all the second class chains have the
following canonical form, in which each constraint $ {}_{(a)}{\bar
\chi}^{(h){'}}_{(k)A^{'}_k}$ has zero Poisson bracket with all the
constraints except with a partner either in its chain or in
another one with the same prefix $(a)$.

\bea
 \begin{array}{llllll}
 {\it pairs\,\, of\,\,  0-chains} & {}& {}_{(1)}{\bar \chi}^{(o){'}}_{(o)A^{'}} &
 {}_{(1)}{\bar \chi}^{(o){'}}_{(o)B^{'}\not= A^{'}} & {} & {\it
 primary}\\
 {} & {} & {} & {} & {} & {}\\
 {\it 1-chains} & {} & {}_{(1)}{\bar \chi}^{(o){'}}_{(1)A^{'}} &
 {} & {} & {\it primary}\\
 {} & {} & {}_{(1)}{\bar \chi}^{(1){'}}_{(1)A^{'}} & {} & {} &
 {\it secondary}\\
 {} & {} & {} & {} & {} & {}\\
 {\it pairs\,\, of\,\, 2-chains} & {} & {}_{(1)}{\bar \chi}^{(o){'}}_{(2)A^{'}}
 & {}_{(2)}{\bar \chi}^{(o){'}}_{(2)B^{'}\not= A^{'}} &
 {} & {\it primary}\\
 {} & {} & {}_{(3)}{\bar \chi}^{(1){'}}_{(2)A^{'}}
 & {}_{(3)}{\bar \chi}^{(1){'}}_{(2)B^{'}\not= A^{'}} & {} & {\it
 secondary}\\
 {} & {} & {}_{(2)}{\bar \chi}^{(2){'}}_{(2)A^{'}} &
 {}_{(1)}{\bar \chi}^{(2){'}}_{(2)B^{'}\not= A^{'}} & {} & {\it
 tertiary}\\
 {} & {} & {} & {} & {} & {}\\
 {\it 3-chains} & {} & {}_{(1)}{\bar \chi}^{(o){'}}_{(3)A^{'}} &
 {} & {} & {\it primary}\\
 {} & {} & {}_{(2)}{\bar \chi}^{(1){'}}_{(3)A^{'}} & {} & {} &
 {\it secondary}\\
 {} & {} & {}_{(2)}{\bar \chi}^{(2){'}}_{(3)A^{'}} & {} & {} &
 {\it tertiary}\\
 {} & {} & {}_{(1)}{\bar \chi}^{(3){'}}_{(3)A^{'}} & {} & {} &
 {\it quaternary}\\
 {} & {} & {} & {} & {} & {}\\
 {\it pairs\,\, of\, 4-chains} & {} & {}_{(1)}{\bar \chi}^{(o){'}}_{(4)A^{'}}
 & {}_{(2)}{\bar \chi}^{(o){'}}_{(4)B^{'}\not= A^{'}} & {} &
 {\it primary}\\
 {} & {} & {}_{(3)}{\bar \chi}^{(1){'}}_{(4)A^{'}} &
 {}_{(4)}{\bar \chi}^{(1){'}}_{(4)B^{'}\not= A^{'}} & {} &
 {\it secondary}\\
 {} & {} & {}_{(5)}{\bar \chi}^{(2){'}}_{(4)A^{'}} &
 {}_{(5)}{\bar \chi}^{(2){'}}_{(4)B^{'}\not= A^{'}} & {} &
 {\it tertiary}\\
 {} & {} & {}_{(4)}{\bar \chi}^{(3){'}}_{(4)A^{'}} &
 {}_{(3)}{\bar \chi}^{(3){'}}_{(4)B^{'}\not= A^{'}} & {} &
 {\it quaternary}\\
 {} & {} & {}_{(2)}{\bar \chi}^{(4){'}}_{(4)A^{'}} &
 {}_{(1)}{\bar \chi}^{(4){'}}_{(4)B^{'}\not= A^{'}} & {} &
 {\it 5-ary}\\
 {} & {} & {} & {} & {} & {}\\
 {\it 5-chains} & {} & {}_{(1)}{\bar \chi}^{(o){'}}_{(5)A^{'}} &
 {} & {} & {\it primary}\\
 {} & {} & {}_{(2)}{\bar \chi}^{(1){'}}_{(5)A^{'}} & {} & {} &
 {\it secondary}\\
 {} & {} & {}_{(3)}{\bar \chi}^{(2){'}}_{(5)A^{'}} & {} & {} &
 {\it tertiary}\\
  {} & {} & {}_{(3)}{\bar \chi}^{(3){'}}_{(5)A^{'}} & {} & {} &
 {\it quaternary}\\
  {} & {} & {}_{(2)}{\bar \chi}^{(4){'}}_{(5)A^{'}} & {} & {} &
 {\it 5-ary}\\
  {} & {} & {}_{(1)}{\bar \chi}^{(5){'}}_{(5)A^{'}} & {} & {} &
 {\it 6-ary},
 \end{array}
 \label{3.25}
 \eea

\noindent and so on with this alternating pattern from even-chains
to odd-chains.

This theorem shows that under suitable regularity conditions on
the singular Lagrangian we cannot obtain neither a chain in which
a primary first class constraints generates a secondary second
class constraint nor two chains, whose primary constraints are a
second class pair and which  generate secondary first class
constraints.

\subsection{Second Class Constraints and Dirac Brackets.}

Second class constraints describe {\it inessential pairs of
canonical variables}, which can be eliminated reducing the number
of the degrees of freedom carrying the dynamics of the singular
system (maybe with the price of a breaking of manifest covariance
and/or of the introduction of non-linearities).

When we have a singular system with the constraint sub-manifold
$\bar \gamma \subset T^*Q$ described by the set ${\tilde {\bar
\zeta}}_{\cal A} = \Big( {\bar \Phi}_{(1){\cal A}_1}, {\bar
\Phi}_{(2){\cal A}_2} \Big)$ of first $\Big( {\bar \Phi}_{(1){\cal
A}_1}\Big)$, ${\cal A}_1=1,..,m-2s_2$, and second $\Big( {\bar
\Phi}_{(2){\cal A}_2}\Big)$, ${\cal A}_2=1,..,2s_2$, class
constraints, we can (at least implicitly) eliminate $s_2$ pairs of
canonical variables with the following procedure. Let us assume
that the second class constraints define a $2(n-s_2)$-dimensional
constraint sub-manifold $\gamma_{(2)}$ of $T^*Q$ containing the
final sub-manifold $\bar \gamma \subset \gamma_{(2)} \subset T^*Q$
and that the regularity conditions are such that the
anti-symmetric matrix

\beq
 {\tilde {\bar C}}_{{\cal A}_2{\cal B}_2} = \Big( \{  {\bar
\Phi}_{(2){\cal A}_2},  {\bar \Phi}_{(2){\cal B}_2} \} \Big),
\label{3.26}
 \eeq

 \noindent which is invertible on $\bar \gamma$ ($det\,
 {\tilde {\bar C}}_{{\cal A}_2{\cal B}_2} {|}_{\bar \gamma} \not=
 0$), is {\it invertible} also on $\gamma_{(2)}$  ($det\,
 {\tilde {\bar C}}_{{\cal A}_2{\cal B}_2} {|}_{\gamma_{(2)}} \not=
 0$) with inverse matrix ${\tilde {\bar C}}^{{\cal A}_2{\cal
 B}_2}$ on $\gamma_{(2)}$. As shown by Dirac \cite{1}, the
 sub-manifold $\gamma_{(2)}$ (in general it is not $T^*Q_{(2)}$
 for some configuration space $Q_{(2)}$) has an induced symplectic
 structure whose Poisson brackets, named {\it Dirac brackets}, are

 \beq
  \{ \bar f, \bar g \}^* = \{ \bar f, \bar g \} - \{ \bar f, {\bar
  \Phi}_{(2){\cal A}_2} \} \, {\tilde {\bar C}}^{{\cal A}_2{\cal
 B}_2} \{  {\bar \Phi}_{(2){\cal A}_2}, \bar g \}.
  \label{3.27}
  \eeq

Besides the standard properties $\{ \bar f, \bar g \}^* =- \{ \bar g, \bar f \}^*$,
$\{ \bar f, {\bar g}_1 {\bar g}_2 \}^* = \{ \bar f, {\bar g}_1
\}^*\, {\bar g}_2 + {\bar g}_1\, \{ \bar f, {\bar g}_2 \}^*$,
\hfill\break $\{ \{ \bar f, \bar g \}^*,\, \bar u \}^* + \{ \{
\bar u, \bar f \}^*,\, \bar g \}^* + \{ \{ \bar g, \bar u \}^*,\,
\bar f \}^* =0$, the Dirac brackets have the extra, easily
verified, properties

\bea
 &&\{ {\bar \Phi}_{(2){\cal A}_2}, \bar f \}^* =0\quad {\it for\,
 every\, \bar f\, and {\cal A}_2}, \, \Rightarrow \quad {\bar
 \Phi}_{(2){\cal A}_2} \equiv 0\quad {\it on\,\, \gamma_{(2)}},
 \nonumber \\
 &&{}\nonumber \\
 &&\{ \bar f, {\bar g}_{(1)} \}^* \approx \{ \bar f, {\bar
 g}_{(1)} \} \quad {\it on\,\, \bar \gamma \subset \gamma_{(2)},\,\,
 for\, {\bar g}_{(1)}\,\, first\, class,\, \bar f\,
 arbitrary},\nonumber \\
 &&{}\nonumber \\
 &&\{ \bar f, \{ {\bar g}_{(1)}, {\bar u}_{(1)} \}^*\, \}^*
 \approx \{ \bar f, \{ {\bar g}_{(1)}, {\bar u}_{(1)} \}\, \}\quad
 {\it on\,\, \bar \gamma \subset \gamma_{(2)}},\nonumber \\
 &&\quad\quad\quad {\it for\, {\bar g}_{(1)}\, and\, {\bar u}_{(1)}
 \,\, first\, class,\, \bar f\, arbitrary}.
 \label{3.28}
 \eea

When we use the Dirac brackets, the Dirac Hamiltonian ${\bar H}_D
= {\bar H}_c^{(F)} + \lambda^{\bar A}(t)\, {\bar \phi}_{\bar A}$
becomes ${\bar H}_D^{'} = {\bar H}_c^{(F){'}} + \lambda^{\bar
A}(t)\, {\bar \phi}_{\bar A}$ with ${\bar H}^{(F){'}}_c = {\bar
H}_c^{(F)} {|}_{\gamma_{(2)}}$. Since the Dirac Hamiltonian is a
first class function, for the Hamilton-Dirac equations we get
 ${{d \bar f}\over {d t}} \cir \{ \bar f, {\bar H}_D \} \approx \{
 \bar f, {\bar H}^{'}_D \}^*\quad {\it on\,\, \bar \gamma \subset
 \gamma_{(2)}}$.

The second class constraints are not generators of canonical
transformations interpretable as Hamiltonian gauge transformations
like first class constraints (see next Section), but, as we shall
see in Section VI, they are the generators of local Noether
extended symmetry transformations under which the singular
Lagrangian has a generalized type of quasi-invariance (instead the
first class constraints generate local Noether symmetry
transformations under which the singular Lagrangian is
quasi-invariant).

\section{Hamiltonian Gauge Transformations and Dirac Obseervables}

In this Section we analyze the Hamiltonian gauge transformations generated by the first-class constraints, the notion of DO and the canonical transformations for the identification of a canonical Darboux basis adapted to both first- and second-class constraints.

\subsection{First Class Constraints and Hamiltonian Gauge Transformations. }

On the final constraint sub-manifold $\bar \gamma \subset T^*Q$
the Dirac Hamiltonian depends on as many arbitrary Dirac
multipliers $\lambda^{\bar A}(t)$ as primary first class
constraints ${\bar \phi}_{\bar A}(q, p) \approx 0$. As a
consequence the solutions $q^i(t)$, $p_i(t)$ of the Hamilton-Dirac
equations are functionals of these arbitrary functions of time and
cannot correspond to measurable observables, which must have a
deterministic dependence on time starting from a given set of
Cauchy initial data.

Let us give the canonical coordinates $q^i_o = q^i(t_o)$, $p_{oi}
= p_i(t_o)$ at time $t_o$: this is interpreted as giving a {\it
physical initial state} for the system. Let us consider the time
evolution of a function $\bar f(q, p)$ from $t_o$ to $t_o + \delta
t$ generated by the Dirac Hamiltonian: $\bar f(q,
p){|}_{t_o+\delta t} \cir \bar f(q, p){|}_{t_o} + \delta t\,\, \{
\bar f(q, p), {\bar H}_D(q, p, \lambda ) \} = \bar f(q,
p){|}_{t_o} + \delta t\,\, \{ \bar f(q, p), {\bar H}_c^{(F)}(q, p)
\} + \delta t\,\, \lambda^{\bar A}(t)\, \{ \bar f(q, p), {\bar
\phi}_{\bar A}(q, p) \}$. If we consider two sets of Dirac
multipliers $\lambda_1^{\bar A}(t)$ and $\lambda_2^{\bar A}(t)$
coinciding at $t_o$ [$\lambda_1^{\bar A}(t_o) = \lambda_2^{\bar
A}(t_o)$], we obtain the result that at time $t_o + \delta t$
there is no uniquely determined value for $q^i(t_o+\delta t)$,
$p_i(t_o+\delta t)$, because for every function we get the
following difference between the two time evolutions
$\triangle_{12}\, \bar f(q, p){|}_{t_o+\delta t} = \delta t\,
 [\lambda_1^{\bar A}(t_o) - \lambda_2^{\bar A}(t_o)]\,\, \{ \bar
 f(q, p), {\bar \phi}_{\bar A}(q, p) \} {|}_{t_o}$.

The only way to recover a deterministic description of the
physical states of the system like in the regular case is to
abandon the assumption that a physical state is uniquely
identified by one and only one set of values of the canonical
coordinates at a given time. {\it In the singular case at each
instant of time many sets of canonical coordinates describe the
same physical state}. Two sets of canonical coordinates whose
difference is $\triangle_{12}\, q^i$, $\triangle_{12}\, p_i$ are
said to be {\it gauge equivalent} and the term $\lambda^{\bar
A}(t)\, {\bar \phi}_{\bar A}(q, p)$ of the Dirac Hamiltonian is
interpreted as the generator of a {\it Hamiltonian gauge
transformation}. Therefore the $m$ primary first class constraints
${\bar \phi}_{\bar A}$ are the generators of the Hamiltonian gauge
transformations responsible of the non-deterministic time
evolution. This means that in the singular case we must:

\noindent i) Find which is the maximal set of Hamiltonian gauge
transformations existing for each given singular system besides
those appearing in the Dirac Hamiltonian (they are the {\it only
ones} allowed in the description of the time evolution).

\noindent ii) Separate the canonical variables in three disjoint
sets:

a) the {Hamiltonian gauge invariant} variables (the so called {\it
Dirac observables}, DO), which have deterministic time evolution so
that a complete set of them at one instant identifies the physical
state of the system at that instant;

b) the {\it inessential} pairs of canonical variables eliminable
by means of the second class constraints with the Dirac brackets;

c) the {\it Hamiltonian gauge variables} which are irrelevant for
the identification of a physical state, because they have an
arbitrary time evolution. The number of gauge variables will
coincide with the number of functionally independent generators of
infinitesimal Hamiltonian gauge transformations, which allow the
reconstruction of their maximal set.

To find the maximal set of infinitesimal Hamiltonian gauge
transformations we shall assume that their generators ${\bar
G}_a(q, p)$, $a=1,..,g$ have the structure of a {\it local
Hamiltonian gauge algebra} $\tilde g$ under the Poisson brackets,
namely $\{ {\bar G}_a(q, p), {\bar G}_b(q, p) \} = {\bar
C}_{ab}{}^c(q, p)\, {\bar G}_c(q, p)$ with some set of {\it
structure functions} ${\bar C}_{ab}{}^c(q, p)$. When the structure functions are constant on
$\bar \gamma$, ${\bar C}_{ab}{}^c = C_{ab}{}^c = const.$, we speak
of a {\it Lie gauge algebra} with structure constants
$C_{ab}{}^c$.

Once the generators ${\bar G}_a$'s are known the next problem is
to define a {\it local Hamiltonian gauge group} ${\cal G}$, i.e.
the {\it finite} Hamiltonian gauge transformations which can be
build with sequences of infinitesimal Hamiltonian gauge
transformations, and then to see whether there exist Hamiltonian
gauge transformations {\it not connected to the identity} ({\it
large} gauge transformations) due to the topological properties of
the system. The Hamiltonian gauge group is said to be {\it local},
because the space of its {\it gauge parameters} (i.e. the
coordinates of its {\it group manifold}) is coordinatized by
arbitrary functions of time $\epsilon^a(t)$, $a=1,..,g$, and not
by numerical constants $\epsilon^a = const.$.

Let us remark that the Hamiltonian gauge transformations are
defined {\it off-shell}, namely without using the equations of
motion, and that in general they are more general of the {\it
standard Lagrangian gauge transformations}. Only {\it on-shell},
namely on the space of the solutions of the equations of motion,
do the two notions of gauge transformations coincide and both of
them are then also {\it gauge dynamical symmetries} of the
equations of motion.

The primary first class constraints ${\bar \phi}_{\bar A}(q, p)$
are in general only a subset of the ${\bar G}_a$'s. Their
associated gauge parameters $\epsilon^{a={\bar A}}(t)$ are the
Dirac multipliers $\lambda^{\bar A}(t) \cir g^{\bar
A}_{(\lambda)}(q, \dot q)$, which identify the non-determined
non-projectable velocity functions of the singular system
associated with the non-invertibility of the equations $p_i =
{\cal P}_i(q, \dot q)$. Therefore there will be an equal number of
{\it primary Hamiltonian gauge variables} ${\bar Q}^{\bar A}(q,
p)$, which are transformed among themselves by the Hamiltonian
gauge transformations generated by the ${\bar \phi}_{\bar A}$'s,
satisfy ${{d {\bar Q}^{\bar A}}\over {d t}} \cir \{ {\bar Q}^{\bar
A}, {\bar H}_D \} = \lambda^{\bar A}(t) \cir g^{\bar
A}_{(\lambda)}(q, \dot q)$ and do not contribute to the
identification of the physical state of the system.

However, the study of the Hamilton-Dirac equations (or of the EL
equations) will show that in general there exist {\it secondary
Hamiltonian (or Lagrangian) gauge variables} ${\bar {\cal
T}}^{\alpha}(q, p)$, which inherit the arbitrariness of the Dirac
multipliers, being functionals of them on the solutions of the
equations of motion. This implies that the gauge parameters
$\epsilon^a(t)$, $a=1,..,g$, of the {\it off-shell} Hamiltonian
gauge transformations have to be restricted to $\epsilon^{a=\bar
A}(t) = \lambda^{\bar A}(t)$, $\epsilon^{a\not= \bar A}(t) =
F^{a\not= \bar A}[\lambda^{\bar B}(t)]$ to be interpretable as the
gauge parameters of the {\it on-shell} Hamiltonian gauge group.

The gauge algebra assumption implies that the Poisson bracket of
two infinitesimal gauge transformations must be a gauge
transformation. Therefore we must have
$\{ {\bar \phi}_{\bar A}(q, p), {\bar \phi}_{\bar B}(q, p) \}
= {\bar C}_{\bar A\bar B}{}^a(q, p)\, {\bar G}_a(q, p)$.

Moreover the gauge algebra must not change in time. This implies
that the time derivative  ${{d {\bar G}_a}\over {d t}}
\cir \{ {\bar G}_a, {\bar H}_D \}$  of a Hamiltonian gauge
transformation must be again a gauge transformation. Since we only
know that $\{ {\bar \phi}_{\bar A}, {\bar \phi}_{\bar B} \} =
{\bar C}_{\bar A\bar B}{}^a\, {\bar G}_a$, we get the following
requirement on the ${\bar \phi}_{\bar A}$'s:
$\{ {\bar \phi}_{\bar A}(q, p), {\bar H}_c^{(F)}(q, p) \} = {\bar
 V}^a_{\bar A}(q, p)\, {\bar G}_a(q, p)$,
where ${\bar H}_c^{(F)}$ is the final canonical
Hamiltonian (a first class quantity).

Let us assume that the regularity conditions on the singular
Lagrangian be such that Theorem 2 of Subsection IIID on the
diagonalization of chains of constraints holds. This means that we
can find linear first class combinations ${\bar
\chi}^{(o)}_{(k){\bar A}_k}$ of the primary constraints such that
$\{ {\bar \chi}^{(o)}_{(k){\bar A}_k}, {\bar H}_c^{(F)} \} = {\bar
\chi}^{(1)}_{(k){\bar A}_k}$. Therefore in this form the secondary
first class constraints are generators of Hamiltonian gauge
transformations. The general result $\{ {\bar
\chi}^{(h)}_{(k){\bar A}_k}, {\bar H}_c^{(F)} \} = {\bar
\chi}^{(h+1)}_{(k){\bar A}_k}$ shows that all the secondary,
tertiary ... first class constraints are generators of Hamiltonian
gauge transformations, namely that ${\bar G}_{\cal A} = {\bar
\Phi}_{(1){\cal A}}$, ${\cal A}=1,..,M$. Since in general $\{
{\bar \Phi}_{(1){\cal A}}, {\bar \Phi}_{(1){\cal B}} \} = {\bar
C}_{{\cal A}{\cal B}}{}^{\cal C}\, {\bar \Phi}_{(1){\cal C}} +
{\bar C}^{'}_{{\cal A}{\cal B}}{}^{{\cal C}^{'}}\, {\bar
\Phi}_{(2){\cal C}^{'}}$, we see that a true gauge algebra is
obtained only near the second class sub-manifold $\gamma_{(2)}$,
which contains the final constraint sub-manifold $\bar \gamma$.

Therefore under suitable regularity conditions on the singular
Lagrangian {\it Dirac's conjecture} \cite{1} that all the first
class constraints are generators of Hamiltonian gauge
transformations is true. Dirac also proposed to replace the final
Dirac Hamiltonian ${\bar H}_D = {\bar H}_c^{(F)} + \lambda^{\bar
A}(t)\, {\bar \phi}_{\bar A}$ with the {\it extended Hamiltonian}

 \beq
 {\bar H}_E = {\bar H}^{(F)}_c + \epsilon^{\cal A}(t)\, {\bar \Phi}_{(1){\cal A}},
 \label{4.1}
 \eeq

\noindent including all the first class constraints each one with
an arbitrary multiplier. In this way the time evolution is split
in a deterministic part governed by the final canonical
Hamiltonian ${\bar H}_c^{(F)}$ (it generates a mapping from a
gauge orbit to another one) and in a gauge part, which is the
generator of the most general {\it off-shell} Hamiltonian gauge
transformation.

Even if this extension does not change the
on-shell dynamics (so that, as shown in Refs.\cite{6,8,9}, it is
taken as the starting point of the BRST quantization program), it
has the drawback that {\it its inverse Legendre transformation does not
reproduce the original singular Lagrangian}, because the secondary
gauge variables have an arbitrary gauge freedom instead of the
reduced one ($\epsilon^{{\cal A}=\bar A}(t) = \lambda^{\bar
A}(t)$, $\epsilon^{{\cal A}\not= \bar A}(t) = F^{{\cal A}\not=
\bar A}[\lambda^{\bar B}(t)]$) associated with the {\it on-shell}
Hamiltonian gauge transformations. Even if they have a reduced
gauge freedom, this is a consequence of the fact that the
secondary gauge variables have non-vanishing Poisson brackets with
the generators ${\bar G}_{\cal A}$. The results: i)  $\{ {\bar
\chi}^{(h)}_{(k){\bar A}_k}, {\bar H}_c^{(F)} \} = {\bar
\chi}^{(h+1)}_{(k){\bar A}_k}$ of Theorem 2; ii) ${\bar H}_c^{(F)}
= {\bar H}_c + {\it (combinations\, of\, primary\, both\, first\,
and\, second\, class\, constraints)}$, imply  that all
the secondary first class constraints ${\bar \chi}^{(h)}_{(k){\bar
A}_k}$, $h \not= 0$, must {\it already be present in the original
canonical Hamiltonian } ${\bar H}_c$ with some form of the primary
${\bar Q}^{\bar A}_{(k)} = {\bar {\cal T}}^{(o){\bar A}_k}_{(k)}$
and secondary $ {\bar {\cal T}}^{(h){\bar A}_k}_{(k)}$, $h > 0$,
gauge variables as coefficients (only the  ${\bar {\cal
T}}^{(f_k){\bar A}_k}_{(k)}$ are not present in ${\bar H}_c$)

\beq
 {\bar H}_c(q, p) = {\bar H}^{'}_c(q, p) + \sum_k\, {\bar Q}^{\bar
 A={\bar A}_k}_{(k)}(q, p)\, {\bar \chi}^{(1)}_{(k){\bar A}_k}(q, p) + \sum_k\,
 \sum_{h=1}^{f_k-1}\,  {\bar {\cal T}}^{(h){\bar A}_k}_{(k)}(q, p)\,
 {\bar \chi}^{(h+1)}_{(k){\bar A}_k}(q, p).
 \label{4.2}
 \eeq

For instance, this is what happens in field theories like
electromagnetism, Yang-Mills theory and metric gravity, which have
the secondary first class constraints already present in the
canonical Hamiltonian density with in front the primary gauge
variables.

Therefore Dirac's proposal (\ref{4.1}) is already fulfilled for
this class of singular Lagrangians, but with the gauge parameters
of the {\it on-shell} Hamiltonian gauge group replacing those of
the {\it off-shell} group present in the extended Hamiltonian.

In Ref.\cite{25} it is shown that also the secondary, tertiary
... second class constraints of this class of singular Lagrangians
are present in the canonical Hamiltonian ${\bar H}_c$ in the form
of quadratic combinations. This is clear if we use the form ${\bar
\chi}^{(h){'}}_{(k)A^{'}_k}$ of the second class constraints given
in Theorem 3. Since these constraints are generated by the
equations $\{ {\bar \chi}^{(h){'}}_{(k) A^{'}_k}, {\bar H}_D^{(F)}
\} = {\bar \chi}^{(h+1){'}}_{(k)A^{'}_k}$, the final form of
${\bar H}_c$ implying these results will be

\bea
 {\bar H}_c(q, p) &=& {\bar H}_d(q, p) + \sum_k\, \Big( {\bar
 Q}^{\bar A={\bar A}_k}_{(k)}(q, p)\, {\bar \chi}^{(1)}_{(k){\bar A}_k}(q, p) +
 \sum_{h=1}^{f_k-1}\, {\bar {\cal T}}^{(h){\bar A}_k}_{(k)}(q, p)\,
 {\bar \chi}^{(h+1)}_{(k){\bar A}_k}(q, p)\Big) +\nonumber \\
 &+& \sum_{k, h_1, h_2}\, {\bar
 S}^{(h_1h_2)A^{'}_kB^{'}_k}_{(k)}(q, p)\, {\bar
 \chi}^{(h_1){'}}_{(k)A^{'}_k}(q, p)\, {\bar
 \chi}^{(h_2){'}}_{(k)B^{'}_k}(q, p).
 \label{4.3}
 \eea

\noindent with suitable functions ${\bar
S}^{(h_1h_2)A^{'}_kB^{'}_k}_{(k)}$ consistent with the pattern of
Poisson brackets of the second class constraints given in Theorem
3. In Eq.(\ref{4.3}) ${\bar H}_d$ is the real first class
deterministic Hamiltonian generating a mapping among the gauge
orbits.

See the bibliography of Refs.\cite{25,29,30} for the attempts to prove Dirac's conjecture and
the use of the extended Hamiltonian. However in many of these papers one uses singular Lagrangians with
a singular Hessian matrix with non constant rank (see Section VI).

\subsection{Dirac Observables, Reduced Phase Space and Gauge Fixings.}

We have found that the generators ${\bar G}_{\cal A}$ of the
maximal set of Hamiltonian gauge transformations are all the first
class constraints ${\bar \Phi}_{(1){\cal A}}$, ${\cal A}=1,..,M$.
Therefore, as already said, from the $2n$ original canonical variables $q^i$, $p_i$
we can form three groups of functions:

\noindent i) $2 s_2$ functions which represent the {\it
inessential degrees of freedom} eliminable by going to Dirac
brackets with respect to the second class constraints ${\bar
\Phi}_{(2){\cal A}^{'}} \approx 0$, ${\cal A}^{'}=1,..,2 s_2$.
They do not determine the physical state of the system, but only
restrict the allowed region of $T^*Q$ to the second class
$2(n-s_2)$-dimensional sub-manifold $\gamma_{(2)}$, on which the
symplectic structure is given by the Dirac brackets, by
eliminating degrees of freedom with trivial first order dynamics.
The Dirac Hamiltonian on $\gamma_{(2)}$ is ${\bar
H}_D{|}_{\gamma_{(2)}} = {\bar H}_d + {\it (terms\, in\, the\,
first\, class\, constraints)}$.

\noindent ii) The non-deterministic $M$ {\it primary and secondary
gauge variables}, which also do not determine the physical state.
Together with the first class constraints ${\bar \Phi}_{(1){\cal
A}} \approx 0$, ${\cal A}=1,..,M$, they form a set of $2 M$
functions not carrying dynamical information (except maybe a
topological one). These constraints determine the final $[2(n-s_2)
- M]$-dimensional sub-manifold $\bar \gamma \subset \gamma_{(2)}
\subset T^*Q$. This sub-manifold, which can be odd-dimensional,
does not admit a symplectic structure (no uniquely defined Poisson
brackets exist for the functions on $\bar \gamma$) and is called a
{\it presymplectic (or co-isotropic) manifold} co-isotropically
embedded in $T^*Q$ \cite{12,13}. When suitable mathematical requirements are
satisfied, it can be shown that the sub-manifold $\bar \gamma$ is {\it foliated}
by $M$-dimensional diffeomorphic leaves, the {\it Hamiltonian gauge orbits}.

\noindent iii) $2(n-M-s_2)$ independent functions ${\bar
F}_{\alpha}(q, p)$ with (in general weakly) zero Poisson bracket
with all the constraints, $\{ {\bar F}_{\alpha}, {\bar
\Phi}_{(1){\cal A}} \} \approx 0$, $\{ {\bar F}_{\alpha}, {\bar
\Phi}_{(2){\cal A}^{'}} \} \approx 0$. They are the gauge
invariant classical Dirac observables (DO) which parametrize the
physical states of the system. The DO's are
those functions on $\bar \gamma$ which are {\it constant on the
gauge orbits}. One DO is the deterministic part
${\bar H}_d$ of the canonical Hamitonian ${\bar H}_c$. If $\bar
F$, $\bar G$ are DO's, then the Jacobi identity
implies that also $\{ \bar F, \bar G \}$ is a DO.
Usually one eliminates the second class constraints by introducing
the associated Dirac brackets $\{ .,. \}^*$ and considering $\bar
\gamma$ a sub-manifold of the second class sub-manifold
$\gamma_{(2)}$.

In the case that the constraint sub-manifold $\bar
\gamma$ is foliated by $M$-dimensional diffeomorphic Hamiltonian
gauge orbits ({\it nice} foliation), we can go to the quotient
with respect to the foliation and define the {\it reduced phase
space} ${\bar \gamma}_R$, which will be a manifold if the
projection $\pi : \bar \gamma \mapsto {\bar \gamma}_R$ is a {\it
submersion}, but in general not a co-tangent bundle $T^*Q_R$ over
some reduced configuration space $Q_R$. In the nice case the
reduced phase space is a symplectic manifold with a closed
symplectic two-form and the Hamiltonian in ${\bar
\gamma}_R$ is the deterministic part ${\bar H}_d$ of the canonical
Hamiltonian and the Hamilton equations for the abstract DO's
are ${{d {\bar F}_R}\over {d t}} \cir \{ {\bar F}_R,
{\bar H}_{dR} \}_R = \{ \bar F, {\bar H}_d \}^*$.

In general things may be much more complicated: there can be not
diffeomorphic gauge orbits, there can be singular points in $\bar
\gamma$, $\,\, \pi : \bar \gamma \mapsto {\bar \gamma}_R$ may not
be a submersion, ... For more details see Ref.\cite{6}.

To avoid technical problems with the definition of the reduced
phase space ${\bar \gamma}_R$, usually we try to build a copy of
it by adding $M$ {\it gauge fixing constraints} ${\bar \rho}_{\cal
A}(q, p) \approx 0$, ${\cal A}=1,..,M$, to eliminate the gauge
freedom by choosing a definite gauge.
The constraints ${\bar \rho}_{\cal A}(q, p) \approx 0$,
${\bar \Phi}_{(1){\cal A}}(q, p) \approx 0$ must form a second class
set and we can define their Dirac brackets. Locally the
hyper-surface ${\bar \rho}_{\cal A}(q, p) \approx 0$ in $T^*Q$
should intersect each Hamiltonian gauge orbit in $\bar \gamma$ in one
and only one point (modulo global problems like the Gribov ambiguity in Yang-Mills theory \cite{6}).

If we use the Dirac Hamiltonian (namely the on-shell
gauge group), we have only $\bar k \leq m$ Dirac multipliers
$\lambda^{\bar A}(t)$ in front of the primary first class
constraints ${\bar \phi}_{\bar A}(q, p) \approx 0$. The procedure
for introducing the gauge fixing constraints in this case has been
delineated in Ref.\cite{31}. Let us use the notation ${\bar
\Phi}_{(1){\cal A}} = \Big( {\bar \chi}^{(h)}_{(k){\bar A}_k}
\Big)$, $h =1,..,f_k$, of Theorem 2 for the constraints. Each
$f_k$-chain of first class constraints starts with the primary
${\bar \chi}^{(o)}_{(k){\bar A}_k}$ [its associated Dirac
multiplier is denoted $\lambda^{{\bar A}_k}_{(k)}(t)$] and ends
with the $(f_k+1)$-ary constraint ${\bar \chi}^{(f_k)}_{(k){\bar
A}_k}$. Let us add as many gauge fixing constraints ${\bar
\rho}^{(f_k)}_{(k){\bar A}_k} \approx 0$ as $(f_k+1)$-ary
constraints. It must be $det\, \Big( \{ {\bar
\rho}^{(f_k)}_{(k){\bar A}_k}, {\bar \chi}^{(f_h)}_{(h){\bar A}_h}
\} \Big) \not= 0$ and, for the sake of simplicity, we assume that
$ \{ {\bar \rho}^{(f_k)}_{(k){\bar A}_k}, {\bar
\chi}^{(o)}_{(h){\bar A}_h} \} = 0$ for all $h$, ${\bar A}_h$. The
requirement that the gauge fixings are preserved in time, i.e.
${{d {\bar \rho}^{(f_k)}_{(k){\bar A}_k}}\over {d t}} \cir \{
  {\bar \rho}^{(f_k)}_{(k){\bar A}_k}, {\bar H}_D \}  =
  {\bar \rho}^{(f_k-1)}_{(k){\bar A}_k} \approx 0$,
generates the gauge fixing constraints ${\bar
\rho}^{(f_k-1)}_{(k){\bar A}_k} \approx 0$ to the $f_k$-ary
constraints ${\bar \chi}^{(f_k-1)}_{(k){\bar A}_k} \approx 0$. The
preservation in time of these induced gauge fixings generates the
gauge fixing constraints for the $(f_k-1)$-ary first class
constraints and so on. Each time we arrive at the gauge fixing of
a primary first class constraint, its preservation in time
determines the Dirac multiplier associated with the chain. In
other words, we first fix the value of the $(f_k+1)$-ary gauge
variables, its time constancy fixes the $f_k$-ary gauge variables
and so on; the preservation in time of the gauge fixing of the
primary gauge variables determines the Dirac multipliers.

Let us remark that due to the difficulties in trying to quantize
the Dirac brackets after the elimination of an arbitrary set of
second class constraints, there have been some attempts to
redefine the theory in such a way that only first class
constraints are present. The gauge fixings to the gauge freedom
associated with these new constraints reproduce the original
theory with its second class constraints. The method
\cite{32,33,34} (see also exercise 1.22 of Ref.\cite{6}) requires an
enlarged phase space with as many new pairs of canonical variables
as pairs of second class constraints. The second class constraints
are transformed into first class ones by inserting a suitable
dependence on the new canonical variables. This method has a great
degree of arbitrariness, modifies the theory off-shell and, having
new gauge invariances, has to redefine the canonical Hamiltonian
and the observables.

\subsection{The Shanmugadhasan Canonical Transformations for the Identification of the Gauge Variables and of the Physical Dirac Observables}

We have defined a DO as a first class function
on phase space restricted to  the constraint sub-manifold $\bar
\gamma$: this means that it must have weakly zero Poisson bracket
with all the first and second class constraints and that, as a
consequence, it is constant on the gauge orbits, namely that it is
associated with a function on the reduced phase space. Since
DO's describe the dynamical content of a singular
dynamical system, it is important to find an algorithm for the
determination of a canonical basis of them to be able to visualize
such a content. This would allow to determine all possible Dirac
observables of the singular system and would open the path to the
attempt to quantize only the dynamical degrees of freedom of the
system as an alternative to Dirac quantization with subsequent
reduction at the quantum level (see for instance the {\it BRST
observables} of the BRST quantization \cite{6}).

This strategy is possible due to the class of canonical
transformations discovered by Shanmugadhasan \cite{35}  studying the reduction to normal
form of a canonical differential system like the EL equations of
singular Lagrangians. By using the Lie theory of {\it function
groups} \cite{36}, Shanmugadhasan showed that
in each neighborhood in $T^*Q$ of a point of the constraint
sub-manifold $\bar \gamma$ there exists local Darboux bases,whose
restriction to $\bar \gamma$ allows one to separate the gauge
variables from a local Darboux basis of DO's . These
canonical transformations are implicitly used by Faddeev and Popov to define the measure of the phase space path integral
and produce a trivialization of the BRST approach.
See also Refs.\cite{37,38}.

Given a $2n$-dimensional phase space $T^*Q$, the set $G$ of all
the functions $\phi ({\bar F}_a)$ of $r$ independent functions
${\bar F}_1(q, p)$,.., ${\bar F}_r(q, p)$ (the basis of $G$) such
that $\{ {\bar F}_a, {\bar F}_b \} = \phi ({\bar F}_c)$ ($a,b,c =
1,..,r$) is said to be a {\it function group of rank} $r$. If
$\phi_1, \phi_2 \in G$, then $\{ \phi_1, \phi_2 \} \in G$. When
$\{ {\bar F}_a, {\bar F}_b \} =0$ for the values of $a$ and $b$,
the function group $G$ is said commutative. A subset of $G$ which
forms a function group is a sub-group of $G$. If two function
groups $G_1$, $G_2$ of rank $r$ have p independent functions in
common, they are the basis of a sub-group of both $G_1$ and $G_2$.
A function $\phi \in G$ is said singular if it has zero Poisson
bracket with all the functions of $G$; the independent singular
functions of $G$ form a sub-group. Given a function group $G$ of
rank $r$, it can be shown that the system of partial differential
equations $\{ {\bar g}, {\bar F}_a \} = 0$, $a=1,..,r$, admits
$2n-r$ independent functions $g_k$, $k=1,..,2n-r$, as solutions
and they define a {\it reciprocal function group} $G^r$ of rank
$2n-r$. The basis functions of $G$ and $G^r$ are in involution
under Poisson brackets.

The following two theorems on function groups and
involutory systems \cite{36} are the basis of Shanmugadhasan
theory:

i) For a non-commutative function group $G$ of rank $r$ there
exists a {\it canonical basis} ${\bar \phi}_1$,.., ${\bar
\phi}_{m+q}$, ${\bar \psi}_1$,.., ${\bar \psi}_m$ with $2m+q=r$
such that

\beq
 \{ {\bar \phi}_{\lambda}, {\bar \phi}_{\mu} \} = \{ {\bar \psi}_{\alpha}, {\bar
 \psi}_{\beta} \} = 0,\qquad \{ {\bar \phi}_{\alpha}, {\bar \psi}_{\lambda} \} =
 \delta_{\alpha\lambda},\quad \alpha ,\beta =1,..,m,\,\, \lambda ,\mu =1,.., m+q.
 \label{4.4}
 \eeq

As a corollary a non-commutative function group $G$ of rank $r$ is
a sub-group of a function group of rank $2n$, whose basis ${\bar
\phi}_1$,.., ${\bar \phi}_n$, ${\bar \psi}_1$,.., ${\bar \psi}_n$
can be chosen so that

\beq
 \{ {\bar \phi}_i, {\bar \phi}_j \} = \{ {\bar \psi}_i, {\bar
 \psi}_j \} = 0,\qquad \{ {\bar \phi}_i, {\bar \psi}_j \} =
 \delta_{ij},\quad i,j=1,..,n.
 \label{4.5}
 \eeq

ii) A system of $2m+q$ independent equations (defining a surface
$\bar \gamma$ of dimension $2(n-m)-q$ in $T^*Q$)

\beq
 {\bar \Omega}_a(q, p) = 0,\qquad a=1,..,2m+q,
 \label{4.6}
 \eeq

\noindent such that $rank\, \{ {\bar \Omega}_a, {\bar \Omega}_b \}
= 2m$, can be substituted by a {\it locally} equivalent system

\beq
 {\bar \phi}_{\lambda}(q, p) = 0,\qquad \lambda =1,..,m+q,\qquad\qquad
 {\bar \psi}_{\alpha}(q, p) = 0, \qquad \alpha =1,..,m,
 \label{4.7}
 \eeq

\noindent for which the relations

\beq
 \{ {\bar \phi}_{\lambda}, {\bar \phi}_{\mu} \} = \{ {\bar
 \psi}_{\alpha}, {\bar \psi}_{\beta} \} =0, \qquad \{ {\bar
 \psi}_{\alpha}, {\bar \phi}_{\lambda}\} = \delta_{\alpha\lambda},
 \label{4.8}
 \eeq

\noindent hold locally in $T^*Q$. Therefore Eqs.(\ref{4.6}) are
equivalent to the vanishing of the canonical basis of a
non-commutative function group of rank $2m+q$.

Let us consider a dynamical system with a $n$-dimensional
configuration space $Q$ described by a singular Lagrangian, whose
associated Hamiltonian description contains: i) a set of first
class constraints ${\bar \Phi}_{(1){\cal A}}(q, p) \approx 0$,
${\cal A} =1,..,M$, among which the primary ones are ${\bar
\phi}_{(o)\bar A}(q, p) \approx 0$, $\bar A=1,..,m$; ii) a set of
second class constraints ${\bar \Phi}_{(2){\cal A}^{'}}(q, p)
\approx 0$, ${\cal A}^{'}=1,..,2s_2$; iii) a final Dirac
Hamiltonian ${\bar H}_D^{(F)} = {\bar H}_c^{(F)}(q, p) +
\sum_{\bar a}\, \lambda^{(o) {\bar A}}(t)\, {\bar \phi}_{\bar
A}(q, p)$. In the $2n$-dimensional phase space $T^*Q$ the dynamics
is restricted to the final $[2(n-s_2)-M]$-dimensional constraint
sub-manifold $\bar \gamma \subset ... \subset \gamma \subset T^*Q$
($\gamma$ is the primary sub-manifold; if there are only first
class constraints $\bar \gamma$ is a presymplectic manifold;
however the term presymplectic manifold is often used to denote a
generic $\bar \gamma$), whose closed degenerate two-form ${\bar
\omega}_{\bar \gamma}$ has $dimension\, ker\, {\bar \omega}_{\bar
\gamma} = M$. We have $\bar \gamma \subset \gamma_{(2)} \subset
T^*Q$, where $\gamma_{(2)}$ is the $2(n-s_2)$-dimensional
sub-manifold defined by the second class constraints with the
symplectic two-form ${\bar \omega}_{(2)}$ giving rise to the Dirac
brackets.

Let the constraints
constraints form a function group of rank $2s_2+m$.

Theorem ii) ensures that in every neighborhood in $T^*Q$ of a
point of $\bar \gamma$ there exists a passive canonical
transformation $( q^i, p_i ) \mapsto ( Q^i, P_i )$ in $T^*Q$ such
that in the new canonical basis the neighborhood is identified by
the new constraints

\beq
 P_{\cal A} \approx 0, \qquad {\cal A} = 1,..,m,\qquad\qquad
 Q^{a^{'}} \approx 0, \quad P_{a^{'}} \approx 0,\qquad a^{'}
 =1,..,s_2.
 \label{4.9}
 \eeq

Therefore locally we obtain an {\it Abelianization} of first class
constraints and a {\it canonical form} of the second class
constraints [$Q^{a^{'}} = {\bar b}^{a^{'}}_{{\cal A}^{'}}\, {\bar
\Phi}_{(2){\cal A}^{'}}$, $P_{a^{'}} = {\bar c}_{a^{'}{\cal
A}^{'}}\, {\bar \Phi}_{(2){\cal A}^{'}}$] associated with this
Abelianization. Eqs.(\ref{4.9}) give the canonical form of a
function group of rank $2s_2+m$. Due to theorem i) the reciprocal
function group of rank $2(n-s_2)-m$ has a basis formed by a) $m$
{\it Abelianized gauge variables} $Q^{\cal A}$ parametrizing the
$m$-dimensional gauge orbits in $\bar \gamma$; b) a canonical
basis of Dirac's observables associated with the Abelianization
described by the $n-m-s_2$ pairs of canonical variables $Q^a$,
$P_a$, $a=1,..,n-m-s_2$, which have zero Poisson brackets with the
constraints in the form (\ref{4.9}) by construction. As a
consequence they have weakly zero Poisson bracket with all the
original constraints, so that they are gauge invariant. This is a
local Darboux basis for the presymplectic sub-manifold $\bar
\gamma$.

Let us remark that the $T^*Q$ Poisson bracket $\{ .,. \}_{Q,P}$
coincides with the Dirac bracket  $\{ .,. \}^*_{\gamma_{(2)}}$
when restricted to the second class sub-manifold $\gamma_{(2)}$.
Therefore it can be shown \cite{37,38} that the new Dirac
Hamiltonian ${\bar H}^{(F){'}}_D$ [$p_i\, dq^i - {\bar
H}^{(F)}_D\, dt = P_i\, dQ^i - {\bar H}^{(F){'}}_D\, dt - dF$] is
the first class function

\bea
 {\bar H}^{(F){'}}_D &=& {\bar H}^{(F){'}}_c(Q, P) + \sum_{\bar a}\,
 \lambda^{(o) {\bar A}}(t)\, {\bar d}_{{\bar A}{\cal B}}\,\, P_{\cal B},\nonumber \\
 &&{}\nonumber \\
  {\bar H}^{(F){'}}_c(Q, P) &=& {\bar K}^{(F)}_c(Q, P) - {\tilde
  {\bar \Phi}}_{(2){\cal A}^{'}}(Q, P)\, {\tilde {\bar c}}_{{\cal
  A}^{'}{\cal B}^{'}}(Q, P)\, \{ {\tilde {\bar \Phi}}_{(2){\cal B}^{'}}(Q,
  P), {\bar K}^{(F)}_c(Q, P) \},\nonumber \\
  &&{\tilde {\bar \Phi}}_{(2){\cal A}^{'}}(Q, P) = {\bar \Phi}_{(2){\cal
  A}^{'}}(q(Q, P), p(Q, P)),\nonumber \\
  && {\tilde {\bar c}}_{{\cal A}^{'}{\cal C}^{'}}\, \{ {\tilde {\bar
  \Phi}}_{(2){\cal C}^{'}}, {\tilde {\bar \Phi}}_{(2){\cal
  B}^{'}}\} = \delta_{{\cal A}^{'}{\cal B}^{'}},\nonumber \\
  &&\{ {\bar K}^{(F)}_c, P_{\cal A} \} = \{ {\bar K}_c^{(F)},
  Q^{a^{'}} \} = \{ {\bar K}_c^{(F)}, P_{a^{'}} \} = 0,
  \label{4.10}
  \eea

\noindent  since we have ${\bar \Phi}_{(1){\cal A}} = {\bar
d}_{{\cal A}{\cal B}}\, P_{\cal B} + (terms\, quadratic\, in\,
the\, second\, class\, constraints) \equiv {\bar d}_{{\cal A}{\cal
B}}\, P_{\cal B}$ near $\bar \gamma$.

When we are able
to solve the first class constraints ${\bar \Phi}_{(1){\cal A}}(q,
p) \approx 0$ in a subset $p_{\cal A}$ of the momenta, as already said in Subsection IIIC, a possible
Abelianized form of the first class constraints is

\beq
 P_{\cal A} = p_{\cal A} - {\bar \psi}_{\cal A}(q^i, p_{i
\not= {\cal B}}) \approx 0.
 \label{4.11}
 \eeq

Therefore with the Shanmugadhasan canonical transformations we are
able to separate the gauge degrees of freedom (either inessential
variables or, in reparametrization invariant theories, variables
describing the generalized inertial effects \cite{4}) from the physical
ones at least {\it locally} in suitable open sets of $T^*Q$
intersecting the constraint sub-manifold $\bar \gamma$.

Moreover we see which kind of freedom we have in the choice of the
functional form of the primary constraints: at least locally we
can always make a choice ensuring the complete diagonalization of
chains discussed in Subsection IIID. In a (in general local)
Shanmugadhasan basis we have

\bea
 &&\{ P_{\cal A}, P_{\cal B} \} = \{ P_{\cal A}, Q^{a^{'}} \} = \{
 P_{\cal A}, P_{a^{'}} \} = 0,\nonumber \\
 &&\{ Q^{a^{'}}, P_{b^{'}} \} = \delta^{a^{'}}_{b^{'}},\qquad \{
 Q^{a^{'}}, Q^{b^{'}} \} = \{ P_{a^{'}}, P_{b^{'}} \} =
 0,\nonumber \\
 &&\{ {\bar H}_d, P_{\cal A} \} = \{ {\bar H}_d, Q^{a^{'}} \} = \{
 {\bar H}_d, P_{a^{'}} \} = 0.
 \label{4.12}
 \eea

An open fundamental problem is the determination of those singular
systems which admit a sub-group of Shanmugadhasan canonical
trnsformations {\it globally defined}  in a neighborhood of the
whole constraint sub-manifold $\bar \gamma$. When this class of
transformations exist, we have a family of {\it privileged}
canonical bases in which the constraint sub-manifold becomes the
direct product of the reduced phase space ${\bar \gamma}_R$ (in
the simplest case ${\bar \gamma}_R = T^*Q_R$ for some reduced
configuration space $Q_R$) by a manifold $\Gamma$ diffeomorphic to
the gauge orbits, $\bar \gamma = {\bar \gamma}_R \times \Gamma$.
When $\bar \gamma$ is a stratified sub-manifold, namely it is the
disjoint union of different strata ${\bar \gamma}_a$ each one with
different standard gauge orbit $\Gamma_a$ (this may happen if the
Hessian matrix has variable rank), the same result may be valid
for each stratum, i.e. ${\bar \gamma}_a = {\bar \gamma}_{R\, a}
\times \Gamma_a$.  The existence of {\it privileged} canonical bases is a
phenomenon induced by the direct product structure and is similar
to the existence of special coordinate systems for the separation
of variables admitted by special partial differential equations.

In general the topological structure of the original configuration
space $Q$ and/or of the constraint sub-manifold $\bar \gamma
\subset T^*Q$ will not allow the existence of this {\it
privileged} class of canonical transformations. For instance this
usually happens when the original configuration space $Q$ is a
{\it compact} manifold. In these cases the only way to study the
constraint sub-manifold is to use the classical BRST cohomological
method \cite{6}. However, even in this cases it is
interesting to extrapolate and define new singular dynamical
systems with this direct product structure from the local results
for the original systems. The study of these new models can give
an idea of the {\it non-topological} part of the dynamics of the
original systems.

Moreover,  special relativity
induces a stratification of the constraint sub-manifold of
relativistic singular systems (all having the Poincare\ group as
the kinematical global Noether symmetry group) according to the
types of Poincare' orbit existing for the allowed configurations
of the singular isolated system. Again each Poincare' stratum has
to be studied separately to see whether it admits the direct
product structure. When such a structure is present the {\it
privileged} canonical bases have to be further restricted by
selecting the ones whose coordinates are also adapted to the
Poincare' group

\section{The Hessian Matrix of Singular Lagrangians and the Euler-Lagrange Equations}

After the description of the Hamiltonian formalism we analyze the EL equations of singular
Lagrangians and their projectability to phase space. Then we give an idea of the pathologies
which can appear when the rank of the Hessian matrix is not constant.

\subsection{The Eigenvalues of the Hessian Matrix and the Classification
of the Euler-Lagrange Equations.}

After having described the first order Hamiltonian formalism for
singular systems, let us come back to the second order formalism
based on the singular Lagrangian $L(q, \dot q)$ and its EL
equations. The singular nature of the system is associated with
the $m \leq n$ null eigenvalues of the $n \times n$ Hessian matrix
$A_{ij}(q, \dot q)$, $det\, \Big( A_{ij}(q,
\dot q) \Big) =0$. This is the source of the $m$ primary
constraints ${\bar \phi}_A(q, p) \approx 0$, $A=1,.,m$, when
$rank\, \Big( A_{ij}(q, \dot q) \Big) \, = n-m =const.$ everywhere
in the $(q, \dot q)$ space.

Since we have $\phi_A(q, {\cal P}(q, \dot q)) = {\bar \phi}_A(q,
p){|}_{p={\cal P}(q,\dot q)} \equiv 0$, we get the identity

\beq
 0 \equiv {{\partial}\over {\partial {\dot q}^i}}\,\,
 \phi_A(q, {\cal P}(q, \dot q)) = A_{ij}(q, \dot q)\,\, {{\partial
 {\bar \phi}_A}\over {\partial p_j}} {|}_{p={\cal P}(q,\dot q)}.
 \label{5.1}
 \eeq

This means that $ {{\partial {\bar \phi}_A}\over {\partial p_i}}
{|}_{p={\cal P}(q,\dot q)}$ is a (non-normalized) {\it null
eigenvector} of the Hessian matrix and that, when the primary
constraints are {\it irreducible}, each choice of their functional
form generates a different basis of $m$ (non-normalized) null
eigenvectors for the $m$-dimensional null eigen-space of
$A_{ij}(q, \dot q)$.

If we saturate the EL equations with the null
eigenvectors $ {{\partial {\bar \phi}_A}\over {\partial p_i}}
{|}_{p={\cal P}(q,\dot q)}$ we get (some of these equations may be
void, $0=0$)

\beq
 {\bar \chi}_A(q,\dot q)\, =  {{\partial {\bar \phi}_A}\over
 {\partial p_i}} {|}_{p={\cal P}(q,\dot q)}\,\, L_i(q, \dot q)\,
 \equiv \,  {{\partial {\bar \phi}_A}\over {\partial p_i}}
 {|}_{p={\cal P}(q,\dot q)}\,\, \alpha_i(q, \dot q)\, \cir 0.
 \label{5.2}
 \eeq

In the singular case the EL equations  are an
autonomous system of ordinary differential equations, which cannot
be put in normal form and which, in general, contains {\it
equations of the second, first and zeroth order} as shown by
non-void equations (\ref{5.2}). The first order EL equations
are then divided in two groups according to whether they either
are or are not projectable to phase space.

\noindent i) The {\it zeroth order EL equations} are those
non-void equations (\ref{5.2}) which depend only on the
configuration coordinates $q^i$'s. They are {\it holonomic
Lagrangian constraints}. Since we always {\it include} among the
configuration variables $q^i$ eventual (linear or non-linear)
Lagrange multipliers, these Lagrangian constraints will appear as
{\it secondary Hamiltonian constraints} ${\bar \chi}^{(1)}(q)
\approx 0$ in $T^*Q$ (the primary Hamiltonian constraint being
given by the vanishing of the canonical momentum of the Lagrange
multiplier).

\noindent ii) The {\it non-projectable first order EL equations}
contained in Eqs.(\ref{5.2}) are the {\it genuine first order
equations of motion}. They are also called the {\it primary SODE
conditions} (see the end of Subsection IIB). By using the extended
Legendre transformation [$g^A_{(\lambda )}(q, \dot q) \mapsto
\lambda^A(t)$ for the canonical form of the velocity functions] we
get that their Hamiltonian version depends on the Dirac
multipliers: in $T^*Q$ these equations are recovered from the {\it
kinematical half} of the Hamilton-Dirac equations on the final
constraint sub-manifold $\bar \gamma$. These genuine first order
equations of motion determine the non-projectable primary (and by
induction also the non-primary) velocity functions associated with
the Hamiltonian second class constraints (see the extended second
Noether theorem in Section VI). Actually they are the counterpart
in the second order formalism of those Hamiltonian equations, like
Eq.(\ref{3.8}), which determine the Dirac multipliers
associated with the primary second class constraints (and
therefore they determine the velocity functions in canonical
form). From Theorems 2 and 3 of Subsection IIID we deduce that the primary SODE
conditions correspond to the determination of the Dirac
multipliers of pairs of second class 0-chains, while the higher
SODE conditions correspond to the determination of the Dirac
multipliers for all the second class 1-, 2- .. chains.

\noindent iii) The {\it projectable first order EL equations}
among Eqs.(\ref{5.2}) are those {\it non-holonomic} (also
said an-holonomic or integrable) {\it Lagrangian constraints}
which are projected to the {\it secondary Hamiltonian constraints}
${\bar \chi}^{(1)}_{a_1}(q, p)$ in $T^*Q$.

\noindent iv) The remaining combinations of the EL equations,
which depend on the accelerations ${\ddot q}^i$ in an essential
way, are the {\it genuine second order equations of motion}.

In addition to the EL equations  there are also their
{\it consequences}, namely all those combinations of the EL
equations and of their time derivatives, which do not depend on
the accelerations. They will form what is called an {\it invariant
system} with respect to the EL equations \cite{35} and can be of
the types i), ii), iii). The consequences of the EL equations of
type ii) are called {\it higher SODE conditions}. These aspects of
the theory of singular systems will be clarified with the second
Noether theorem in Section VI.

Let us remark that the presence of first order EL equations implies that
the first order velocity space formalism in the tangent bundle $TQ$ (described
in Subsection IIC in the regular case) cannot be extended to singular systems and, as
said in Refs.\cite{23,24,25}, till now there is no working formulation but only by hand extensions.

\subsection{Pathologies of Singular Systems with Hessians of Variable Rank}

As shown in Ref \cite{24} (where all the examples quoted in the reported bibliography are analyzed and clarified) when the Hessian matrix has not a constant rank
many types of pathologies may appear. To put control on them the basic point is to look for a Hamiltonian formulation of these systems implying
that the Euler-Lagrange equations and the Hamilton equations have the same solutions.

The main pathologies are :

\noindent $\alpha$) {\it Third and Fourth Class Constraints.}

These new types of non-primary constraints are at the basis of the
failure of Dirac's conjecture for many singular Lagrangians with a
Hessian matrix of variable rank. This happens because these
constraints $\bar \chi (q, p) \approx 0$ look like first class
constraints. Their associated Hamiltonian vector fields ${\bar
X}_{\bar \chi} = \{ ., \bar \chi \}$ are either first class,
namely tangent to the constraint sub-manifold $\bar \gamma$, or
vanishing on $\bar \gamma$ (but not near $\bar \gamma$). However,
they are {\it not generators of Hamiltonian gauge transformations}. Instead, in general they generate
{spurious solutions of the Jacobi equations}, which are
not deviations between two neighboring solutions of the EL
equations, due to the linearization instability present in these
singular systems. As an example consider a non-primary constraint $p_1 \approx 0$: i) if it is
first class, its conjugate variable $q^1$ is a gauge variable; ii) if it is second
class there is another constraint determining $q^1$, so that the pair $q^1, p_1$ can be eliminated; iii)
if it is {\it third class}, the
conjugate variable $q^1$ is {\it determined by one combination of
the final Hamilton-Dirac equations} and depends on the initial data.

Instead a {\it fourth class or ineffective constraint}  is a non-primary
constraint $\bar \chi (q, p) \approx 0$ generated inside a chain
by the Dirac algorithm such that $d \bar \chi {|}_{\bar \chi =0}
=0$ even if all the other constraints in the chain have a
non-vanishing differential. These constraints have weakly
vanishing Poisson brackets with every function on $T^*Q$, so that
they are first class quantities. For the sake of simplicity, let
us consider $p^2_1 \approx 0$ as a non-primary constraint of this
type. For the determination of the constraint sub-manifold $\bar
\gamma$ we must use its linearized form $p_1 \approx 0$. But we
cannot use this linearized form in the final Dirac Hamiltonian
generating the final Hamilton-Dirac equations (as instead is
usually done), because otherwise the solutions of the
Hamilton-Dirac and EL equations do not coincide.

\noindent $\beta$) {\it Proliferation of Constraints and
Ramification of Chains of Constraints.}

Let us consider the chains of constraints discussed in Subsection IIID. If in a chain
one gets a constraint like $q^1\, q^2 \approx 0$ (this possible only if the Hessian rank
is not constant) then the chain gives rise to three distinct chains
(ramification of chains) because the constraint
gives rise to the following three sectors: i) $q^1 \approx 0$, $q^2 \approx 0$ (proliferation of
constraints); ii) $q^1 \approx 0$, $q^2 \not= 0$; iii) $q^2 \approx 0$, $q^1 \not= 0$.

\noindent $\gamma$) {\it Joining of Chains of Constraints.}

In certain examples after some steps after a ramification of chains there could be a joining of two
of the new chains.

\bigskip

Look at Ref. \cite{24} for all the examples of these pathologies and for what is known in
mathematical physics on singular systems. Even if we discard all the pathological cases with
Hessians of constant rank, there is not a consistent formulation of singular systems covering
the second order formalism, the tangent space one and the cotangent Hamiltonian one.

\section{Singular Lagrangians and the Second Noether Theorem for Finite-Dimensional Systems}

After quoting the first Noether theorem and its extensions, we show that behind singular Lagrangians
and Hamiltonian constraints there is an extension of the second Noether theorem.

\subsection{Symmetries, the First Noether Theorem and its Extensions.}

The two Noether theorems are a basic ingredient in the
study of the consequences of the invariances of Lagrangian systems
under continuous symmetry transformations. Ref.\cite{39} gives a
review of their applications in theoretical physics, while
Ref.\cite{40} contains a review of the intrinsic geometrical
formulations and of the various extensions of the first Noether
theorem and Refs.\cite{41,42,43} survey the use of the second theorem.
In this Subsection we shall review the first theorem and its
extensions.

For finite-dimensional systems described by a regular (maybe time
dependent) Lagrangian $L(t, q, \dot q)$ the first Noether theorem
states that if the action functional $S = \int dt\, L$ is {\it
quasi-invariant} under a $r$ parameter group $G_r$ of continuous
transformations of $t$ and $q^i$, then $r$ linear independent
combinations of the EL equations $L_i$ reduce identically to total
time derivatives. The converse is also true under appropriate
hypotheses.

This means that if under an infinitesimal set of invertible {\it
local}  variations $\delta_at = {\bar t}_a - t = \delta_a t(t,
q)$, $\delta_{oa} q^i = {\bar q}^i_a(t) - q^i(t) = \delta_{oa}
q^i(t, q, \dot q)$ \footnote{The associated {\it global}
variations (corresponding to Lie derivatives) are $\delta_a q^i =
{\bar q}^i_a ({\bar t}_a) - q^i(t) = \delta_{oa} q^i + {\dot
q}^i\, \delta_at$. The corresponding variations of the velocities
are $\delta_{oa} {\dot q}^i = {d\over {d t}}\, \delta_{oa}q^i$,
$\delta_a {\dot q}^i = \delta_{oa} {\dot q}^i - {\dot q}^i\, {{d
\delta_at}\over {d t}}$.}, $a=1,..,r$,
the total variation of $L$ is a total time derivative (the
following equation is a Killing-type equation; $\equiv$ means {\it
identically} )

\bea
 \delta_a L &=& L\Big( {\bar t}_a, {\bar q}_a({\bar t}_a), {{d
 {\bar q}_a({\bar t}_a)}\over {d {\bar t}_a}} \Big)\, {{d {\bar
 t}_a}\over {d t}} - L(t, q, \dot q) =\nonumber \\
 &=& {{\partial L}\over {\partial q^i}}\, \delta_{oa} q^i +
 {{\partial L}\over {\partial {\dot q}^i}}\, \delta_{oa} {\dot
 q}^i + {d\over {d t}} ( L\, \delta_a t ) =\nonumber \\
 &=& {{\partial L}\over {\partial t}} \, \delta_a t +
 {{\partial L}\over {\partial q^i}}\, \delta_a q^i +
  {{\partial L}\over {\partial {\dot q}^i}}\, \delta_a {\dot
 q}^i + L\, {{d \delta_a t}\over {d t}} =\nonumber \\
 &=& \delta_{oa} q^i\, L_i + {d\over {d t}} \Big(
 {{\partial L}\over {\partial {\dot q}^i}}\, \delta_{oa} {\dot
 q}^i + L\, \delta_a t \Big) \equiv {{d F_a(t, q, \dot q)}\over
 {d t}},
 \label{6.1}
 \eea

\noindent then one obtains the following $r$ {\it Noether
identities} (both $G_a$ and $F_a$ are in general functions of
 $t$, $q^i$ and ${\dot q}^i$)

\bea
 {{d G_a}\over {d t}} &\equiv& - \delta_{oa} q^i\, L_i \cir 0,
 \nonumber \\
 {}&& \nonumber \\
 G_a &=& {{\partial L}\over {\partial {\dot q}^i}}\,
 \delta_{oa} q^i - F_a + L \delta_a t = {{\partial L}\over
 {\partial {\dot q}^i}}\, \delta_a q^i - F_a -\Big( {\dot q}^i\,
 {{\partial L}\over {\partial {\dot q}^i}} - L\Big) \delta_a t.
 \label{6.2}
 \eea

The $r$ quantities $G_a(q, \dot q)$ are {\it constants of the
motion} (in field theory one would obtain $r$ conservation laws
$\partial_{\mu}\, J^{\mu}_a \cir 0$). For $F_a \not= 0$ we speak
of {\it quasi-invariance}, while for $F_a =0$ of {\it invariance}.
When we have $\delta_{oa} q^i(t, q)$, we get $F_a(t, q)$.

It is always possible to define a new set of variations in which
$t$ is not varied ($\delta^{'}_a t =0$, $\delta^{'}_{oa} q^i =
\delta_{oa} q^i$) and which gives rise to the same constants of
motion $G_a$: the only difference is that now $\delta^{'}_a L
\equiv {{d F^{'}_a}\over {d t}}$ with $F^{'}_a = F_a - L \delta_a
t$. In general, there is an infinite family of Noether symmetry
transformations $\delta_a t$, $\delta_{oa} q^i$ associated with
the same set of constants of motion $G_a$ (even a change of the
functional form of the Lagrangian is allowed: $\delta_a L =
L^{'}(barred\,\, variables)- L$). See Ref.\cite{40} for a critical review and the
proposal of a preferred geometrical approach.
Moreover, inside every family of Noether symmetry transformations
there are always {\it dynamical symmetry} transformations, i.e. symmetry transformations of the EL differential
equations mapping the space of its solutions onto itself (the sets
of Noether symmetry and dynamical symmetry transformations of a
Lagrangian system do not coincide but have an overlap).

The concept of a family of Noether transformations associated with
a given set of constants of motion has also been analyzed by
Candotti, Palmieri and Vitale \cite{44}. They point out that each
family contains transformations ${\tilde \delta}_a t$, ${\tilde
\delta}_{oa} q^i$ such that Eqs.(\ref{6.1}) become
${\tilde \delta}_a L \equiv {{d {\tilde F}_a}\over {d t}} +
 f_a,\quad\quad f_a(t, q, \dot q, \ddot q) \cir 0$.
That is we have a {\it weak} quasi-invariance, because $\delta_a
L$ only becomes a total time derivative by using the EL equations.
The Noether identities (\ref{6.2}) become
${{d {\tilde G}_a}\over {d t}} \equiv - {\tilde \delta}_{oa} q^i\,
 L_i + f_a \cir 0$ and give rise to the same constants of motion $G_a$, if ${\tilde \delta}_a t$, ${\tilde
\delta}_{oa} q^i$, ${\tilde F}_a$ are such that ${\tilde G}_a =
{{\partial L}\over {\partial {\dot q}^i}}\, {\tilde \delta}_{oa}
q^i - {\tilde F}_a + L {\tilde \delta}_a t \cir G_a$. In the
regular case these extensions can be considered irrelevant, but it
is not so in the singular case.

\bigskip

The {\it generator} of the Noether transformation is the vector
field $Y = \delta t\, {{\partial}\over {\partial t}} + \delta q^i\,
{{\partial}\over {\partial q^i}} + \delta {\dot q}^i\, {{\partial}\over {\partial {\dot
 q}^i}}$ and in terms of it we get $\delta t = Y t$, $\delta q^i
= Y q^i = {\cal L}_Y \, q^i$, $Y L \equiv \dot F - L\, {{d \delta
t}\over {d t}}$. The constant of motion $G = {{\partial L}\over {\partial {\dot
q}^i}}\, \delta_o q^i - F + L\, \delta t$ is  an {\it
invariant} of the generator $Y$: $Y\, G = 0$.

The natural setting for the definition and study of the generator
$Y$ (and of the dynamical symmetries of differential equations) is
the {\it infinite jet bundle} \cite{45}, where $Y$ is a {\it
Lie-B\"acklund vector field} ($Y$ gives its
truncation to the first derivatives). As shown in Refs.\cite{45}
there are only two kinds of invariance transformations when the
number of degrees of freedom is higher than one: i) the Lie point
tranformations of $R \times Q$ extended to the higher derivatives;
ii) the Lie-B\"acklund transformations (or {\it tangent
transformations of infinite order} preserving the tangency of
infinite order of two curves). These latter have $\delta t$ and/or
$\delta_o q^i$ depending on the velocities and possibly on the
higher accelerations. For instance, the non-point canonical
transformations of $T^*Q$ become Lie-B\"acklund transformations
with $\delta t =0$, when rephrased through the inverse Legendre
transformation in the second order Lagrangian formalism.

\bigskip

By expressing the velocities in terms of the coordinates and
canonical momenta we get  $\delta q^i(q,\dot q) = {\bar {\delta
q^i}}(q,p)$, $F(q, \dot q) = \bar F(q,p)$, $G(q, \dot q) = \bar
G(q,p) = p_i\,  {\bar {\delta q^i}}(q,p) - \bar F(q,p)$. Since the
Hamiltonian is defined as $\bar H = p_i\, {\dot q}^i - L$, the
phase space Lagrangian satisfies $\bar L(q,p,\dot q) = p_i\, {\dot
q}^i - \bar H(q,p) = L(q, \dot q)$ and therefore will have the
same invariance properties. This means $\delta \bar L = {\dot q}^i\, {\bar {\delta p_i}} + p_i {{d {\bar {\delta
 q^i}}} \over {d t}} - {{\partial \bar H}\over {\partial q^i}}\,
 {\bar {\delta q^i}} - {{\partial \bar H}\over {\partial p_i}}\,
 {\bar {\delta p_i}} = {\bar {\delta p_i}}\, {\bar L}_{qi} - {\bar {\delta q^i}}\,
 {\bar L}_{pi} + {d \over {d t}} ( p_i \, {\bar {\delta q^i}} )
 \equiv {{d \bar F}\over {d t}}$. Therefore we get
${d \over {d t}}\, \bar G = {d \over {d t}}\, (p_i\, {\bar {\delta q^i}}
 -\bar F) \equiv - {\bar {\delta p_i}}\, {\bar L}_{qi} + {\bar {\delta
 q^i}}\, {\bar L}_{pi} \cir 0,\quad \Rightarrow \, \{ \bar G, \bar H \} \cir 0$
 and ${\bar {\delta q^i}} \equiv {{\partial \bar G}\over {\partial
 p_i}} = \{ q^i, \bar G \}$,
 ${\bar {\delta p_i}} \equiv - {{\partial \bar G}\over {\partial
 q^i}} = \{ p_i, \bar G \}$. This is the phase space projected Noether identity associated to
the constant of motion $\bar G$.

In this way we have found that the Hamiltonian Noether symmetry
transformation is generated by the constant of motion $\bar G(q,
p)$ ($\{ \bar G, \bar H \} =0$) considered as the generator of a
symmetry canonical transformation, i.e. such that the functional
form of the Hamiltonian does not change ($\delta_o\, H = {\bar
H}^{'}(q, p) - \bar H(q, p) =0$).

The intrinsic formulation of the first Noether theorem and of the
reduction of dynamical systems with symmetry, when there is a free
and proper symplectic action of a (connected) Lie group on a
(connected) symplectic manifold (phase space of an autonomous
regular Hamiltonian system with symmetry), is the momentum map
approach \cite{46}. For a {\it weakly regular value} of the
momentum map associated with this action  the reduced phase space
has a structure of symplectic manifold and inherits  a Hamiltonian
dynamics. For a singular value of the momentum map the reduced
phase space is a {\it stratified symplectic manifold} \cite{47}.

\subsection{The Second Noether Theorem in the Second Order
Lagrangian Formalism and its Extension.}

The second Noether theorem states that if the action functional
$S\, =\, \int dt L$ is {\it quasi-invariant} (i.e. its variation
is a total time derivative) with respect to an infinite continuous
group $G_{\infty r}$, involving up to order $k$ derivatives (i.e.
a group whose general transformations depend upon $r$ essential
arbitrary functions $\epsilon^a(t)$ and their first $k$ time
derivatives), then $r$ {\it identities} exist among the EL
equations $L_i$ and their time derivatives up to order $k$. Under
appropriate hypotheses the converse also is true.

This version of the theorem is oriented to the description of
gauge theories and general relativity, in which there is a
singular Lagrangian invariant under local gauge transformations
and/or space-time diffeomorphisms and giving rise  {\it only to
first class} constraints at the Hamiltonian level (see for instance Ref.\cite{48}).
This is unsatisfactory, because at the Lagrangian level the
fundamental property of singular Lagrangians is the number of null
eigenvalues of the Hessian matrix and some of them may be
associated with Hamiltonian second class constraints (when
present). As a consequence in the literature there is no clear
statement about the connection between the second Noether theorem
and the canonical transformations generated by the constraints
when some of them are second class \cite{7}. An {\it extension}
of the second Noether theorem is needed to include these cases.
This was done in Ref.\cite{49} along the lines of the
Candotti-Palmieri-Vitale extension \cite{45} of the first Noether
theorem using the concept of {\it weak quasi-invariance} in the case of a Hessian matrix of constant rank.

The {\it extended second Noether theorem} may be expressed by
saying that the action functional associated with a singular
Lagrangian is {\it weakly} quasi-invariant (i.e. quasi-invariant
only after having used combinations of the EL equations and of
their time derivatives which are independent of the accelerations)
under as many sets of local infinitesimal Noether transformations
as is the number of null eigenvalues of the Hessian matrix. Each
set of such transformations $\delta_A\, q^i$ depends on an
arbitrary function $\epsilon^A(t)$ and its time derivatives up to
order $J_A$ and produces an identity which can be resolved in $J_A
+ 2$ {\it Noether identities}, each one being the time derivative
of the previous one.

While the $\delta_A\, q^i$'s associated with chains of first class
constraints will turn out to be the pull-back by means of the
inverse Legendre transformation of the {\it infinitesimal
Hamiltonian gauge canonical transformations}, the $\delta_A\,
q^i$'s associated with chains of second class constraints will
turn out to be the pull-back of the infinitesimal canonical
transformations generated by the second class constraints (they
could be named {\it pseudo-gauge transformations}).

This local formulation of the extended theorem which is based on a form of
the infinitesimal Noether transformations $\delta_A\, q^i$ valid
if we use the orthonormal eigenvectors ${}_A{\hat \xi}^i_o(q, \dot
q) = \tau^i_A(q, \dot q) = {{\partial {\hat {\bar \phi}}_A(q,
p)}\over {\partial p_i}} {|}_{p={\cal P}(q, \dot q)} $ of the
Hessian matrix. Their use corresponds to the
diagonalization of the chains of constraints discussed in Subsection
IIID and allows to show that the $J_A + 2$ Noether identities of the form of Eqs.(\ref{6.2}) implied by
the generalized weak quasi-invariance are projectable to phase
space, where they rebuild the whole Dirac algorithm (each chain of identities is connected with
a chain in the theorems 2 and 3 of Subsection IIID). See also the next Section.

\section{Constraints in Field Theories }

The naive extension of the previous results to classical field
theory does not present conceptual problems
\cite{1,5,6,7,50,51}. Instead a more rigorous
treatment would require much more sophisticated techniques: see
for instance Ref.\cite{18} for an introduction to infinite
dimensional Hamiltonian systems. The new real phenomenon of field
theory with constraints is the possible appearance of the zero
modes of the elliptic operators associated with some constraints
(due to the spatial gradients of the fields and/or the canonical
momenta). It depends on the choice of the function space for the
fields, an argument on which there is no general consensus, and
creates obstructions to the existence of global gauge fixing
constraints like the Gribov ambiguity in Yang-Mills theory
(its existence depends upon the choice of the function space \cite{52}).

Let us suppose that we have a singular Lagrangian density ${\cal
L}(\varphi^r(x), \varphi^r_{,\mu}(x))$  depending upon a set of
fields $\varphi^r(x)$, $r=1,..,n$ and their first derivatives
$\varphi^r_{,\mu}(x) = \partial_{\mu}\, \varphi^r(x)$. The
space-time manifold $M$ of dimension $m+1$ (usually the
4-dimensional Minkowski space-time) has local Cartesian
coordinates $x^{\mu} = ( x^o, x^i)$ and Lorentzian metric
$\eta^{\mu\nu} = (1; -1,..,-1)$. The action is the local
functional $S = \int d^{m+1}x\, {\cal L}$ and a certain class of
boundary conditions at $|\vec x| \rightarrow \infty$ for the
fields has been chosen in some function space dictated by physical
considerations.

With the usual non-covariant choice of $x^o$ as time variable we have the following
definition of Hessian matrix

\beq
 A_{rs}(x)\, {\buildrel {def}\over =}\, A^{oo}_{rs}(\varphi^s(x),
 \varphi^s_{,\mu}(x)) = {{\partial^2 {\cal L}}\over {\partial
 \varphi^r_{,o}(x)\, \partial \varphi^s_{,o}(x)}},\qquad
 A^{\mu\nu}_{rs} = {{\partial^2 {\cal L}}\over {\partial
 \varphi^r_{,\mu}\, \partial \varphi^s_{,\nu}}}.
 \label{7.1}
 \eeq

The EL equations implied by Hamilton's principle $\delta\, S =
\delta\, \int_{\Omega}\, d^{m+1}x\, {\cal L} = 0$ for arbitrary
variations $\delta\, \varphi^r(x)$ vanishing on the boundary
$\partial \Omega$ of the compact region $\Omega$  are

 \bea
 L_r(x) &=& {{\partial {\cal L}}\over {\partial \varphi^r(x)}} -
 \partial_{\mu}\, {{\partial {\cal L}}\over
 {\partial\varphi^r_{,\mu}(x)}} = - ( A^{\mu\nu}_{rs}(x)\,
 \varphi^s_{,\nu}(x) -\alpha_r(x)) = - (A_{rs}(x)\,
 \varphi^s_{,oo}(x) - {\hat \alpha}_r(x)),\nonumber \\
 &&{}\nonumber \\
 &&\alpha_r - {{\partial {\cal L}}\over {\partial \varphi^r}} -
 {{\partial^2 {\cal L}}\over {\partial \varphi^r_{,\mu}\, \partial
 \varphi^s}}\, \varphi^s_{,\mu},\qquad {\hat \alpha}_r = \alpha_r
 - 2\, A^{oi}_{rs}\, \varphi^s_{,oi} - A^{ij}_{rs}\,
 \varphi^s_{,ij}.
 \label{7.2}
 \eea

The canonical momenta and the standard Poisson brackets (after a
suitable definition of functional derivative)  are

\beq
 \pi_r(x) = \Pi^o_r(x),\qquad \Pi^{\mu}_r(x) = {{\partial {\cal
 L}}\over {\partial \varphi^r_{,\mu}(x)}},\qquad
 \{ \varphi^r(x^o, \vec x), \pi_s(x^o, \vec y) \} = \delta^r_s\,
 \delta^m(\vec x - \vec y).
 \label{7.3}
 \eeq

Since the Poisson brackets are {\it local} (i.e. they do not
depend on primitives of the delta function, a property named {\it
local commutativity} in Ref.\cite{6}) the Poisson bracket of two
local functionals is also a local functional, so that non-local
terms cannot be generated through the operation of taking the
bracket. For two functions ${\bar F}_a(\varphi^r(x^o, \vec x),
\varphi^r_{,i}(x^o, \vec x), \pi_r(x^o, \vec x), \pi_{r\, ,i}(x^o,
\vec x))$, $a=1,2$, we have

\beq
 \{ {\bar F}_1(x^o, \vec x), {\bar F}_2(x^o, \vec y) \} = \int
 d^{m+1}z\, \Big[ {{\delta {\bar F}_1(x^o, \vec x)}\over {\delta
 \varphi^r(x^o, \vec z)}}\, {{\delta {\bar F}_2(x^o, \vec x)}\over {\delta
 \pi_r(x^o, \vec z)}} - {{\delta {\bar F}_1(x^o, \vec x)}\over {\delta
 \pi_r(x^o, \vec z)}}\, {{\delta {\bar F}_2(x^o, \vec x)}\over {\delta
 \varphi^r(x^o, \vec z)}} \Big].
 \label{7.4}
 \eeq

 Since $det\, A_{rs}(x) = 0$, there will be a certain number of
 null eigenvalues of the Hessian matrix with associated local
 orthonormal null eigenvectors $\tau^r_A(\varphi^s(x),
 \varphi^s_{,\nu}(x))$, $A=1,..,n_1$. For the sake of simplicity we
 shall assume regularity conditions such that the Hessian matrix
 has a constant rank everywhere.

As in the finite-dimensional case there are $n_1$ arbitrary
velocity functions non projectable to phase space and $n_1$
primary constraints ${\bar \phi}_A(\varphi^r, \varphi^r_{,i},
\pi_r, \pi_{r,i}) \approx 0$ such that ${\bar \phi}_A(\varphi^r,
\varphi^r_{,i}, \pi_r, \pi_{r,i}) {|}_{\pi_r={\cal P}_r(\varphi ,
\varphi_{,\mu})} \equiv 0$. Again we have that ${{\partial {\bar
\phi}_A}\over {\partial \pi_r}}$ are null eigenvalues of the
Hessian matrix. For the sake of simplicity we assume that there is
a global functional form of the constraints producing the
orthonormal eigenvectors ${}_A\xi^i_o$.

If the Lagrangian density is sufficiently regular that the
Legendre transformation is well defined, the canonical Hamiltonian
density is ${\bar {\cal H}}_c(\varphi^r(x), \varphi^r_{,i}(x),
\pi_r(x), \pi_{r,i}(x)) = \varphi^r_{,o}(x)\, \pi_r(x) - {\cal
L}(\varphi (x), \varphi_{,\mu}(x))$, while the Dirac Hamiltonian
is

\beq
 {\bar H}_D = \int d^mx\, \Big( {\bar {\cal H}}_c(x^o, \vec x) +
 \sum_A\, \lambda^A(x^o, \vec x)\, {\bar \phi}_A(x^o, \vec x) \Big) = {\bar
 H}_c + \sum_A\, {\bar H}_A.
 \label{7.5}
 \eeq

 Only by choosing consistent boundary conditions for the fields
 and the Dirac multipliers we can interpret the ${\bar H}_A$'s as
 generators of local Noether transformations. The Hamilton
 equations are

 \bea
  {\bar L}^r_{D \varphi} &=& \varphi^r_{,o}(x^o, \vec x) - \{
  \varphi^r(x^o, \vec x), {\bar H}_D \} \cir 0,\nonumber \\
  {\bar L}_{Dpr} &=& p_{r,o}(x^o, \vec x) - \{ p_r(x^o, \vec x),
  {\bar H}_D \} \cir 0.
  \label{7.6}
  \eea

Having the primary constraints ${\bar \phi}_A(x^o, \vec x) \approx
0$ and the Dirac Hamiltonian Dirac's algorithm is plainly extended
to field theory starting with the study of the time constancy of
the primary constraints. One arrives at a final constraint
sub-manifold $\bar \gamma$, divides the final set of constraints
in first and second class ones and determines the Dirac
Hamiltonian of $\bar \gamma$ as ${\bar H}_D = {\bar H}_c + \int
d^mx\,  \sum_{\bar A}\, \lambda^{\bar A}(x^o, \vec x)\, {\bar
\phi}_{\bar A}(x^o, \vec x)$, where ${\bar \phi}_{\bar A}(x^o,
\vec x) \approx 0$ are the first class primary constraints and
$\lambda^{\bar A}(x^o, \vec x)$ the Dirac multipliers, equal to
the primary arbitrary velocity functions $g^{\bar A}_{(\lambda
)}(\varphi^r(x^o, \vec x), \varphi^r_{,\mu}(x^o, \vec x))$ through
the first half of the Hamilton-Dirac equations.

However, besides regularity conditions on the singular Lagrangian
density  ${\cal L}$ so to avoid the (non-explored) field theory
counterparts of the pathologies of Subsection VB, one has to
consider extra requirements peculiar to field theory:

i) The constraints must define a sub-manifold of the
infinite-dimensional phase space, whose properties depend on the
choice of the boundary conditions and the function space for the
fields and their canonical momenta. This function space must
include all physically interesting solutions of the Hamilton-Dirac
equations. The constraints must not only be local functionals of
the fields but must also be {\it locally complete} \cite{6}. This
means that every phase space function vanishing on $\bar \gamma$
is zero by virtue of the constraints defining $\bar \gamma$ and
their {\it spatial} derivatives of any order {\it only, without
having to invoke the boundary conditions}. To put mathematical control on BRST cohomology
(Ref.\cite{6}, theorem 12.4) one needs strong regularity
conditions implying that every function vanishing on $\bar \gamma$
can be written as combination of the constraints and a finite
arbitrary number of their spatial derivatives.

ii) In Ref.\cite{53} it is pointed out that in field theory each
constraint $\bar \phi (x^o, \vec x) \approx 0$ represents a
continuous and infinite number of constraints characterized by the
space label $\vec x$ so that problems may arise with the theory of
distributions. Subtle difficulties may appear  in the division of
the constraints in the first and second class groups and in the
mathematical definition of Dirac brackets where the inverse of
continuous matrices $C(x^o, \vec x, \vec y)$ are needed.

A related problem is with the gauge transformations generated by
first class constraints $\bar \phi (x^o, \vec x) \approx 0$. If we
consider the most general generator $\bar G = \int\, d^mx\, \alpha
(x^o, \vec x)\, {\bar \phi}(x^o, \vec x)$, is $\bar G$ a generator
of gauge transformations for every parameter function $\alpha
(x^o, \vec x)$? In Ref.\cite{54} the following distinction
between {\it proper} and {\it improper} gauge transformations was
given:

a) Proper gauge transformations represent true gauge symmetries of
the theory and do not change the physical state of the system.
They can be eliminated by fixing the gauge.

b) Improper gauge transformations (they do not exist for
finite-dimensional systems) do change the physical state of the
system, mapping (on-shell) one physical solution onto a different
physical solution. They cannot be eliminated by fixing the gauge
but only by means of {\it superselection rules} selecting a
particular set of solutions.

Given the function space ${\cal F}$ for the fields, the problem is
the determination of the allowed function space of the parameter
functions $\alpha (x^o, \vec x)$ so that $\bar G$ is the generator
of a proper gauge transformation. Not to over-count the
constraints the space ${\cal F}_d$ of the allowed $\alpha (x^o,
\vec x)$ (the so-called {\it dual space}) must be such that, when
$\alpha$ varies in this space, ${\bar G} \approx 0$ has an
information equivalent to the original constraints $\bar \phi
(x^o, \vec x) \approx 0$. If $\alpha (x^o, \vec x)$ does not
belong to the dual space, then $\bar G$ is the generator of an
improper gauge transformation. Under both proper and improper
gauge transformations a field belonging to ${\cal F}$ must be
transformed in a field still in ${\cal F}$. Therefore, if $\varphi
(x^o, \vec x) \in {\cal F}$ and $\pi (x^o, \vec x) \in {\cal F}$
are the fields, then $\delta\, \varphi (x^o, \vec x) = \{ \varphi
(x^o, \vec x), {\bar G} \} = {{\delta \bar G}\over {\delta \pi
(x^o, \vec x)}}$ and $\delta\, \pi (x^o, \vec x) = \{ \pi (x^o,
\vec x), {\bar G} \} = {{\delta \bar G}\over {\delta \varphi (x^o,
\vec x)}}$ must be in ${\cal F}$. Moreover the functional
derivatives must be well defined, namely we must have $\delta\,
\bar G = \int d^mx\, \Big(  {{\delta \bar G}\over {\delta \varphi
(x^o, \vec x)}}\, \delta\, \varphi (x^o, \vec x) + {{\delta \bar
G}\over {\delta \pi (x^o, \vec x)}}\, \delta\, \pi (x^o, \vec x)$.
In general to get this result one has to do a number of
integrations by parts and to check whether the resulting surface
terms vanish: if they vanish we have a proper gauge transformation
with $\alpha (x^o, \vec x) \in {\cal F}_d$. If the $\alpha (x^o,
\vec x)$ are such that the surface terms do not vanish, we have to
modify the generator $\bar G$ by adding a surface term, $\bar G
\mapsto {\bar G}^{'} = \bar G + {\bar G}_{ST}$, whose variation
$\delta\, {\bar G}_{ST}$ cancels the unwanted surface terms: in
this case ${\bar G}^{'}$ is the generator of an improper gauge
transformation and ${\bar G}^{'} \not= 0$ on $\bar \gamma$, where
it becomes the constant surface term ${\bar G}_{ST}$ commuting
with the Hamiltonian. This constant surface term is a non-trivial
constant of the motion which can be fixed only with a
super-selection rule.

iii) In Ref.\cite{55} it is pointed out that
the study of the formal integrability of the partial differential
Hamilton equations requires the use of prolongation methods in the
infinite jet bundle (namely we have to consider derivatives of the
original equations till the needed order) and in particular the
determination of a {\it system of equations in involution}. While
Dirac's algorithm considers all possible consequences of taking
the time derivatives of the Hamilton equations, it says nothing
about their spatial derivatives. Therefore in field theory one has
to check whether extra integrability conditions appear by
considering these spatial derivatives.

\bigskip

In the regular case the first Noether theorem implies the existence of conservation laws
$\partial_{\mu}\, G^{\mu}(x) \cir 0$, so that with suitable boundary conditions on the fields
conserved charges $Q = \int d^mx\, G^o(x^o, \vec x)$, ${{d Q}\over {d x^o}} \cir 0$ are obtained.

In the singular case, by using the orthonormal eigenvectors of the
Hessian matrix the extended {\it second Noether theorem} states
that with each null eigenvalue of this matrix is associated a
local Noether transformation $\delta_A\, x^{\mu} = 0$,
 $\delta_A\, \varphi^r(x) = \epsilon^A(x)\,\, {}_A\xi^r_{J_A}(x)
 + \sum_{j=1}^{J_A}\, \epsilon^A{}_{,\mu_1...\mu_j}(x)\, \,
 {}_A\xi^{r\, (\mu_1...\mu_j)}_{J_A-j}(x)$ ($(\mu_1...\mu_j)$ means symmetrization in the
indices), under which we get the following {\it weak
quasi-invariance}

\beq
 \delta_A\, {\cal L} = \delta_A\, \varphi^r\, L_r +
 \partial_{\mu}\, \Big( {{\partial
 {\cal L}}\over {\partial \varphi^r_{,\mu}}}\, \delta_A\, \varphi^r\Big)
 \equiv \partial_{\mu}\, F^{\mu}_A + \epsilon^A(x)\, D_A \cir
 \partial_{\mu}\, F^{\mu}_A.
 \label{7.7}
 \eeq

Here $F^{\mu}_A = F^{\mu}_A(\varphi , \varphi_{,\nu}, \epsilon^A)$
and $D_A(\varphi , \varphi_{,\nu}) \cir 0$ by using the
acceleration-independent consequences of the EL equations. By
posing

\bea
 F^{\mu}_A(\varphi , \varphi_{,\nu}, \epsilon^A) &=&
 \epsilon^A(x)\,\, {}_AF^{\mu}_{J_A}(\varphi , \varphi_{,\nu}) +
 \sum_{j=1}^{J_A}\, \epsilon^A{}_{,\mu_!...\mu_j}(x)\,\,
 {}_AF^{\mu\, (\mu_1...\mu_j)}_{J_A-j}(\varphi ,
 \varphi_{,\nu}),\nonumber \\
 G^{\mu}_A(\varphi , \varphi_{,\nu}, \epsilon^A) &=& {{\partial
 {\cal L}}\over {\partial \varphi^r_{,\mu}}}\, \delta_A\,
 \varphi^r - F^{\mu}_A = \epsilon^A(x)\,\, {}_AG_{J_A}^{\mu} +
 \sum_{j=1}^{J_A}\, \epsilon^A{}_{,\mu_!...\mu_j}(x)\,\,
 {}_AG^{\mu\, (\mu_1...\mu_j)}_{J_A-j}(\varphi ,
 \varphi_{,\nu}),\nonumber \\
 &&{}_AG^{\mu\, (\mu_1...\mu_j)}_{J_A-j} = {{\partial {\cal
 L}}\over {\partial \varphi^r_{,\mu}}}\,\, {}_A\xi^{r\,
 (\mu_1...\mu_j)}_{J_A-j} - {}_AF^{\mu\, (\mu_1...\mu_j)}_{J_A-j},
 \label{7.8}
 \eea

\noindent we get the following Noether identities

\beq
 \partial_{\mu}\, G^{\mu}_A \equiv \epsilon^A(x)\, D_A -
 \delta_A\, \varphi^r\,\, L_r \cir 0,
 \label{7.9}
 \eeq

\noindent which imply

\bea
 {}_AG_o^{(\mu_1\, (\mu_2...\mu_{J_A+1}))} &\equiv& 0,\nonumber \\
 \partial_{\mu}\,\, {}_AG_o^{\mu\, (\mu_1...\mu_{J_A})} &\equiv& -
 {}_AG_1^{(\mu_1\, (\mu_2...\mu_{J_A}))} - {}_A\xi_o^{r\,
 (\mu_1...\mu_{J_A})}\,\, L_r,\nonumber \\
 &&....\nonumber \\
 \partial_{\mu}\,\, {}_AG_{J_A-j}^{\mu\, (\mu_1...\mu_j)} &\equiv&
 - {}_AG_{J_A-j+1}^{(\mu_1\, (\mu_2...\mu_j))} - {}_A\xi^{r\,
 (\mu_1...\mu_j)}_{J_A-j}\,\, L_r,\quad j=1,..,J_A-1,\nonumber \\
 &&...\nonumber \\
 \partial_{\mu}\,\, {}_AG_{J_A}^{\mu} \equiv D_A -
 {}_A\xi^r_{J_A}\,\, L_r.
 \label{7.10}
 \eea

These equations imply the following form of the Noether identities

\bea
 \partial_{\mu_1}\, ...\, \partial_{\mu_{J_A+1}}\,\,
 {}_AG_o^{\mu_1\, (\mu_2...\mu_{J_A+1}))} &\equiv& 0,\nonumber \\
 &&...\nonumber \\
  \partial_{\mu_1}\, ...\, \partial_{\mu_{j+1}}\,\,
 {}_AG_{J_A-j}^{\mu_1\, (\mu_2...\mu_{j+1}))} &\equiv&
 \sum_{h=0}^{J_A-j-1}\, (-)^{J_A-j-h}\,\, \partial_{\mu_1}\, ...\,
 \partial_{\mu_{J_A-h}}\, \Big( {}_A\xi^{r\,
 (\mu_1...\mu_{J_A-h})}_h\,\, L_r \Big) \cir 0,\nonumber \\
 &&.....\qquad j=1,..,J_A,\nonumber \\
 D_A &\equiv& \sum_{h=0}^{J_A}\, (-)^{J_A-h}\,\,
 \partial_{\mu_1}\, ...\, \partial_{\mu_{J_A-h}}\, \Big(
 {}_A\xi^{r\, (\mu_1...\mu_{J_A-h})}\,\, L_r \Big) \cir 0.
 \label{7.11}
 \eea

When we have $D_A \equiv 0$ (quasi-invariance, first class
constraints) we get the {\it contracted Bianchi identities}

\beq
 \sum_{h=0}^{J_A}\, (-)^{J_A-h}\,\,
 \partial_{\mu_1}\, ...\, \partial_{\mu_{J_A-h}}\, \Big(
 {}_A\xi^{r\, (\mu_1...\mu_{J_A-h})}\,\, L_r \Big) \equiv 0.
 \label{7.12}
 \eeq

We have given a formulation of the theorem based only on the
variation of the Lagrangian density. Usually, in absence of second
class constraints, namely with $D_A(x) \equiv 0$, and with
$\delta_A\, \varphi^r$ depending only on $\epsilon^A(x)$ and
$\partial_{\mu}\, \epsilon^A(x)$, one considers the variation of
the action evaluated on a compact region $\Omega$ of the
$m$-dimensional space bounded by two hyper-planes ($\Sigma_f$ at
$x^o_f$ and $\Sigma_i$ at $x^o_i$; the variations are assumed to
vanish on the spatial boundary) and asks for $\delta\, S = 0$.
Then the identity (\ref{7.9}) becomes $\int_{\Omega}\,
d^{m+1}x\, \delta_A\, \varphi^r\,\, L_r \equiv \int_{\Sigma_i} \,
d^m\sigma_{\mu}\, G^{\mu}_A(x^o_i, \vec x) - \int_{\Sigma_f}\,
d^m\sigma_{\mu}\, G^{\mu}_A(x^o_f, \vec x)$ [$d^m\sigma_{\mu} =
d^mx\, n_{\mu}$ with $n_{\mu}$ outer normal to the hyper-plane],
namely it take the form of a term in the interior of $\Omega$
equated to boundary terms on the hyper-planes. If we ask that the
arbitrary functions $\epsilon^A(x)$ and their derivatives vanish
on the boundary, $\delta_A\, S = 0$ (original Noether statement)
implies the vanishing of the interior term: $\delta_A\,
\varphi^r\, L_r \equiv 0$ and this gives the contracted Bianchi
identities. By combining this result with the identities one can
recover Utiyama \cite{56} and Trautman\cite{57} results, i.e. their form
of the identities. For a detailed discussion of this point and of
the connected interpretative ambiguities  see
Refs.\cite{58,59,60}.

If in the Noether identity (\ref{7.9}) we put $\epsilon^A(x) =
const.$, this global sub-group of gauge transformations gives rise
to the first Noether theorem associated with the Noether
transformation $\delta_A\, \varphi^r = \epsilon^A\,\,
{}_A\xi^r_{J_A}$ as a sub-case of the second theorem and to the
Noether identity

\beq
 \partial_{\mu}\,\, {}_AG^{\mu}_{J_A} \equiv D_A -
 {}_A\xi^r_{J_A}\,\, L_r \cir 0.
 \label{7.13}
 \eeq

Eqs.(\ref{7.13}) are the {\it weak conservation laws} and the
{\it weak} (so called) {\it improper conserved (Noether) current}
is ${}_AG^{\mu}_{J_A}$. Instead it can be checked that the {\it
strong conservation laws} $\partial_{\mu}\, V^{\mu}_A \equiv 0$
hold independently from the EL equations for the
following {\it strong improper conserved current} (it is not a
Noether current)

\beq
 V^{\mu}_A = {}_AG^{\mu}_{J_A} - \sum_{h=0}^{J_A-1}\,
 (-)^{J_A-h}\,\, \partial_{\mu_1}\, ...\,
 \partial_{\mu_{J_A-h-1}}\, \Big( {}_A\xi_h^{r\,
 (\mu\mu_1...\mu_{J_A-h-1})}\,\, L_r \Big) = \partial_{\nu}\, U^{[\mu\nu ]}_A
 \cir {}_AG^{\mu}_{J_A},
 \label{7.14}
 \eeq

\noindent where $U^{[\mu\nu ]}_A$ ($[\mu\nu ]$ means
antisymmetrization) is the following {\it super-potential}

\beq
 U^{[\mu\nu ]}_A = \sum_{h=1}^{J_A-1}\, (-)^{J_A-h}\,\,
 \partial_{\mu_1}\, ...\, \partial_{\mu_{J_A-h-1}}\, \Big(
 {}_AG_h^{(\nu\, (\mu\mu_1...\mu_{J_A-h-1}))} -
 {}_AG_h^{(\mu\, (\nu\mu_1...\mu_{J_A-h-1}))} \Big).
 \label{7.15}
 \eeq

The {\it improper strong $Q_A^{(S)}$ and weak $Q_A^{(W)}$
conserved charges} (${{d Q_A^{(W)}}\over {d x^o}} \cir 0$, ${{d
Q_A^{(S)}}\over {d x^o}} \equiv 0$ for suitable boundary
conditions) coincide on the solutions of the
acceleration-independent EL equations [$\Omega$ is a spatial
volume with boundary $\partial \Omega$]

\beq
 Q_A^{(S)} = \int_{\Omega}\, d^mx\, V^o_A(x^o, \vec x) =
 \int_{\partial \Omega}\, d^{m-1}\Sigma_k\,\, U_A^{[ok]}(x^o, \vec
 x)\,\,\, \cir\,\,\, Q_A^{(W)} = \int_{\Omega}\, d^mx\, \, {}_AG^o_{J_A}(x^o, \vec
 x).
 \label{7.16}
 \eeq

 The source of the ambiguities \cite{58,59,60} is this doubling
 of the conserved currents and charges, which does not exist with
 the first Noether theorem applied to global symmetries.

Finally the {\it generalized Trautman strong conservation laws}
are (differently from Eqs.(\ref{7.9}) they hold independently
from the EL equations)

\beq
 \partial_{\mu}\, \Big[ G^{\mu}_A - \sum_{h=0}^{J_A-1}\,
 \epsilon^A{}_{,\mu_1...\mu_h}(x)\, \sum_{j=h}^{J_A}\,
 (-)^{j-h}\,\, \partial_{\mu_{h+1}}\, ...\, \partial_{\mu_j}\,
 \Big( {}_A\xi_{J_A-j}^{r\, (\mu\mu_1...\mu_j)}\,\, L_r \Big)
 \Big] \equiv 0,
 \label{7.17}
 \eeq

\noindent where for $j=h$ the derivatives acting on the round
bracket are absent.

Let us remark (see Ref.\cite{60}) that when a  field theory has
global symmetry quasi-invariances  the first Noether theorem
implies the existence of a current which is conserved by using the
EL equations: it is the analogue of the weak current, while there
is no analogue of the strong current which exists only with gauge
symmetries. This gives rise to a conserved charge (the analogue of
the weak charge; there is no strong charge in the form of the flux
through the surface at infinity of some vector field) and the
possibility of a symmetry reduction of the order of the system of
equations of motion. Instead in the case of local gauge symmetries
we get super-selection rules, not symmetry reduction.

The Noether identities (\ref{7.10}) and (\ref{7.11}) have not
yet been studied in detail like in the finite-dimensional case,
because they contain a lot of information which is not really
needed for the Hamiltonian treatment.

\section{Final Remarks}

The problem of quantization of systems with constraints is completely open. The standard approach is
the BRST-BV approach  for which we send to  Refs.\cite{6,61}. Besides the ordering problem which may
create inequivalent quantum systems associated to different orderings of the constraints not quadratic in the
canonical variables, there is the problem whether the algebra of quantum constraints based on commutators
remains of the same type (first-, second-class or both) as with Poisson brackets.

See Ref.\cite{11} for an approach to quantization oriented to loop quantum gravity and Ref.\cite{62} for the polymer quantization method.

Another ambiguity is whether we quantize all the variables and then we eliminate the gauge ones at the quantum level
like it happens with the BRST approach, or whether we eliminate the gauge variables at the classical level and we quantize
only the physical degrees of freedom \cite{25,61}. The quantization should be independent from the choice of gauge!

In Ref.\cite{25,63}
there is a multi-temporal formulation in which every gauge variable is re-interpreted as a "time", with the suggestion that the
quantization should be independent from both the ordinary time and the generalized ones.

\vfill\eject

\end{document}